\documentclass[conference]{IEEEtran}
\AtBeginDocument{%
  }
\let\labelindent\relax
\usepackage{epsfig,endnotes}     
\usepackage{colortbl}
\usepackage{url}                 
\usepackage{amsmath}
\usepackage{enumitem}
\usepackage{algorithm}
\usepackage{algpseudocode}
\usepackage{booktabs}
\usepackage{makecell}
\usepackage{float}
\usepackage{multirow}
\usepackage{multicol}
\usepackage{pifont}
\usepackage{microtype}
\usepackage{fancybox}
\usepackage{diagbox}
\usepackage[dvipsnames]{xcolor}

\newif\ifcomment

\newcommand{\xxu}[1]{\ifcomment{\color{blue}
\emph{[XXU: #1]}}\fi}

\newcommand{\rd}[1]{\ifcomment {\color{purple}
\emph{[Ruyi: #1]}} \fi}
\newcommand{\nbo}[1]{{\sf\color{orange}[#1]}}
\newcommand{\lls}[1]{\ifcomment{\nbo{LS: #1}}\fi}

\newcommand{\yf}[1]{\ifcomment{\color{red}
\emph{[Fei: #1]}}\fi}

\newif\ifupdate
\updatetrue
\newcommand{\update}[1]{\ifupdate{\color{black}{#1}}\fi}

\newif\ifcomments

\usepackage{xspace}
\newcommand{\MyMethod}{EncoderLock\xspace}
\microtypecontext{spacing=nonfrench}

\DeclareMathOperator{\argmax}{argmax}
\algrenewcommand\algorithmicrequire{\textbf{Input:}}
\algrenewcommand\algorithmicensure{\textbf{Output:}}

\definecolor{koblue}{HTML}{f74d4d}
\definecolor{kopink}{HTML}{0c84c6}
\definecolor{darkblue}{HTML}{002c53}
\definecolor{darkpink}{HTML}{d7a29c}
\definecolor{target}{HTML}{bf7334}
\definecolor{source}{HTML}{388498}
\definecolor{src-domain}{HTML}{98ad8b}
\definecolor{boxcolor}{HTML}{f2f2f2}
\usepackage{tikz}

\begin{document}
\title{Probe-Me-Not: Protecting Pre-trained Encoders \\ from Malicious Probing}
\author{\IEEEauthorblockN{Ruyi Ding, Tong Zhou,  Lili Su, Aidong Adam Ding, Xiaolin Xu, Yunsi Fei}
        \IEEEauthorblockA{Northeastern University, Boston, MA 02115, USA\\
        \{ding.ruy, zhou.tong1, l.su, a.ding, x.xu, y.fei\}@northeastern.edu
        }
        }
\maketitle

\begin{abstract}
Adapting pre-trained deep learning models to customized tasks has become a popular choice for developers to cope with limited computational resources and data volume.
More specifically, probing--training a downstream head on a pre-trained encoder--has been widely adopted in transfer learning, which helps to prevent overfitting and catastrophic forgetting.
However, such generalizability of pre-trained encoders raises concerns about the potential misuse of probing for harmful intentions, such as discriminatory speculation and warfare applications. 
In this work, we introduce EncoderLock, a novel applicability authorization method designed to protect pre-trained encoders from malicious probing, i.e., yielding poor performance on specified prohibited domains while maintaining their utility in authorized ones.
Achieving this balance is challenging because of the opposite optimization objectives and the variety of downstream heads that adversaries can utilize adaptively.
To address these challenges, EncoderLock employs two techniques: \textit{domain-aware weight selection and updating} 
to restrict applications on prohibited domains/tasks, and \textit{self-challenging training scheme} that iteratively strengthens resistance against any potential downstream classifiers that adversaries may apply.
Moreover, recognizing the potential lack of data from prohibited domains in practical scenarios, we introduce three EncoderLock variants with different levels of data accessibility: \textit{supervised} (prohibited domain data with labels), \textit{unsupervised} (prohibited domain data without labels), and \textit{zero-shot} (no data or labels available).
Extensive experiments across fifteen domains and three model architectures demonstrate EncoderLock's effectiveness over baseline methods using non-transferable learning. 
Additionally, we verify EncoderLock's effectiveness and practicality with a real-world pre-trained Vision Transformer (ViT) encoder from Facebook.
These results underscore the valuable contributions EncoderLock brings to the development of responsible AI.
\end{abstract}
\newcommand{\cmark}{\ding{51}}%
\newcommand{\xmark}{\ding{55}}%

\setlength{\fboxsep}{1pt} 

\section{Introduction} \label{sec: introduction}
As the complexity of learning tasks increases, leveraging pre-trained models becomes a popular strategy for developers to train their customized models efficiently.
Among various transfer learning methods, model probing has emerged as one of the most common and lightweight strategies to utilize pre-learned knowledge effectively~\cite{belinkov2022probing, cao2022putting, white2021non}.
It involves freezing the encoder parts of pre-trained models while fine-tuning only the downstream heads. 
The encoders often include early layers of pre-trained models with more complex structures, which is responsible for extracting useful information from raw data to latent representations, on which downstream heads perform specific tasks such as classification and generation~\cite{he2022masked, le2020contrastive}.

Probing offers several advantages, including resource efficiency, because of its low requirements on data and computational resources, and semantic consistency, as it helps avoid catastrophic forgetting--the performance reduction due to the encoder's loss of pre-learned knowledge after extensive fine-tuning~\cite{chen2019catastrophic, chronopoulou2019embarrassingly, iman2023review}. 
Furthermore, probing allows the pre-trained encoder to be used as a black-box, either as local private models~\cite{dong2023puma, gupta2023sigma, liu2021secdeep} or cloud services through APIs~\cite{clarifai, openai, qu2023reaas}, ensuring better intellectual property protection~\cite{tan2022federated}. 
\update{Nowadays, many companies, such as Clarifai~\cite{clarifai} and OpenAI~\cite{openai}, offer commercial encoder APIs, allowing users to input data and obtain latent feature vectors, which can then be used for various downstream real-world applications.}

However, the general availability of the pre-trained encoder for probing also raises concerns about \textbf{\textit{malicious probing}}, i.e., users can probe the encoder for unethical or harmful tasks~\cite{cheatham2019confronting}.
Examples include building classification heads for discriminatory speculation~\cite{internationalwomensday_gender_ai, whitehouse_ai_bill_of_rights} or autonomous weapons in warfare applications~\cite{marr2018artificial}. 
To address these concerns, model owners have set strict policies regarding the utilization of pre-trained encoders.
For instance, OpenAI\footnote{\url{https://openai.com/policies/terms-of-use}} explicitly prohibits users from employing their encoder services for ``any illegal, harmful, or abusive activity".
However, relying solely on policies, without concrete technological barriers, is insufficient to prevent model misuse. 
Considering malicious probing not only poses ethical risks but also represents a serious form of infringement on the intellectual property of model owners, 
design-time countermeasures are urgently needed for protecting the encoders with applicability authorization~\cite{wang2022non}. 

Proactively preventing pre-trained encoders from malicious probing presents three challenges.
\textbf{Challenge 1: \textit{Integrity of Pre-trained Encoder.}} 
The protection strategy should maintain the encoder's functionality on \textit{authorized domains} (those for which the encoder is designed), while restricting misuse on \textit{prohibited domains} (those not allowed due to malicious intent). 
Furthermore, it is advisable to have a small impact on \textit{admissible domains} (those are gray-listed and not explicitly considered during encoder design).
\textbf{Challenge 2: \textit{Robustness to Malicious Probing.}} Malicious users can customize downstream heads with various configurations (e.g., hyper-parameters and classifier architectures). 
The protection method must be robust against these diverse setups.
\textbf{Challenge 3: \textit{Accessibility to Prohibited Domains.}} Effective protection requires pre-defined prohibited domains, while a lack of samples from these domains can significantly impact its performance.
A few studies design protection against direct applicability on prohibited tasks--malicious users can do inference but no further fine-tuning~\cite{wang2023model, wang2022non}. 
They introduced a training strategy considering solely a given prohibited dataset with clear labels and an authorized dataset, called {\em Non-Transferable Learning (NTL)}.
Unfortunately, NTL doesn't apply to pre-trained encoders--as malicious users can further probe encoders with downstream heads using prohibited data.
Therefore, we propose \textbf{\MyMethod}, a new applicability authorization strategy for pre-trained encoders.

\begin{figure}[t]
    \centering
    \includegraphics[width=0.95\linewidth]{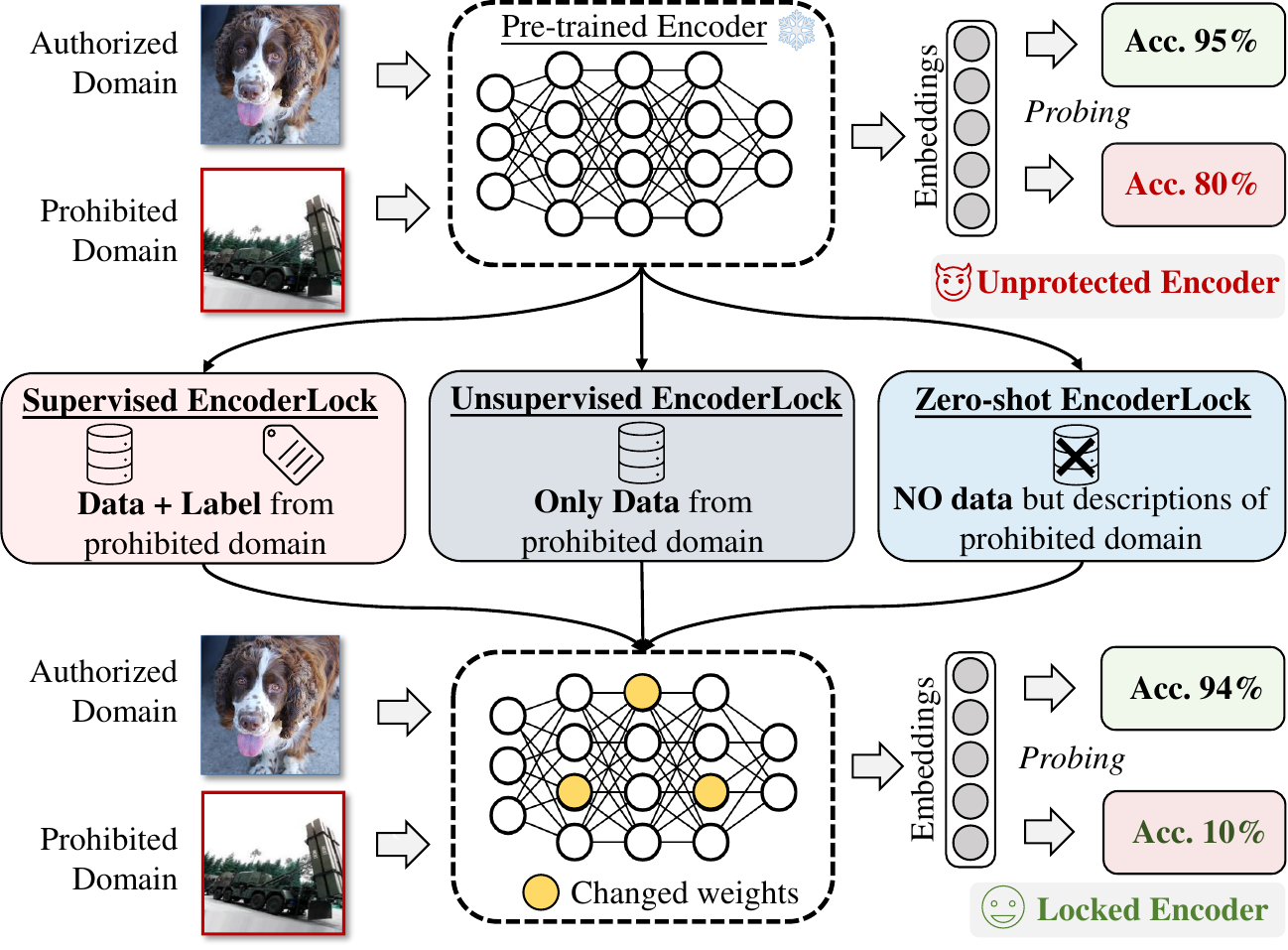}
    \caption{\textbf{Applicability Authorization with \MyMethod}: Fixed pre-trained encoders accept user inputs and return representations. Users can utilize them for various customized tasks by probing with downstream heads. \MyMethod aims to prevent malicious probing to pre-defined prohibited domains, which may have different levels of data accessibility, marked by different colors.}
    \label{fig: motivation-example}
\end{figure}

\MyMethod is based on our new three-level threat model for model applicability authorization for pre-trained encoders, following a paradigm akin to that used in representation learning, as illustrated in Fig.~\ref{fig: motivation-example}: a) \colorbox{koblue!20}{Level 1 Label-enriched}: The provider has a labeled dataset of the prohibited domain,
b) \colorbox{darkblue!20}{Level 2 Label-free}: The provider only has an unlabeled dataset,
c) \colorbox{kopink!20}{Level 3 Theme-only}: The provider has no data but knows the theme they wish to exclude the encoder from processing.
\yf{semantic seems to be a too broad term to use here} \rd{agreed.}
These levels represent real-world model providers with different data accessibility. Throughout the paper, we will use this color coding to represent these data accessibility levels.

\MyMethod proposes the following solutions to address all three challenges against malicious probing.
First, we propose \textit{domain-aware weight selection and updating}, which identifies critical weights to the target domain and adjusts them, successfully restricting the model's transferability to the target domain while minimizing its effect on other authorized domains (addressing Challenge 1). 
To ensure the robustness of \MyMethod against customized malicious downstream heads (addressing Challenge 2), we introduce a minimax optimization--\textit{self-challenging training}, which refines the encoder's feature space iteratively by continuously adjusting auxiliary downstream heads.
Together, these strategies constitute  \colorbox{koblue!20}{\textit{supervised \MyMethod}}, which effectively addresses the Level 1 scenario.
To address Challenge 3, we extend two \MyMethod variants for stricter accessibility to the prohibited domain.
For Level 2 where only an unlabeled target dataset is available, we introduce \colorbox{darkblue!20}{\textit{unsupervised \MyMethod}}, including a novel regularization term based on contrastive loss in the feature space, which deliberately obfuscates features in the target dataset.
For Level 3, we propose \colorbox{kopink!20}{\textit{zero-shot \MyMethod}}, which leverages an AI agent and a text-to-image generative model to build a reliable pathway from semantic description to an unlabeled synthetic dataset. 
To ensure the synthetic dataset is representative of the target domain and comprehensive, we propose a prompt refining method utilizing the AI agent. 

\textbf{Our Contributions:} We propose \textbf{\MyMethod}, a novel and proactive protection on the pre-trained encoder against malicious probing. The contributions of this work include:
\begin{enumerate}[leftmargin=*, itemsep=0pt, topsep=0.5pt]
\item \MyMethod provides a robust applicability authorization framework to owners of pre-trained encoders. It maintains the encoder's performance on authorized domains with the domain-aware weight selection algorithm and offers robust defense against diverse customized probing through a self-challenging training scheme.
\item We propose a three-level threat model following the practical data availability of representation learning. Correspondingly, we present three variants of \MyMethod with novel techniques to address different levels of target domain data accessibility, tackling realistic comprehensive scenarios.
\item We conduct extensive experiments to evaluate \MyMethod across twelve domains and three encoder architectures, including a large, real-world Vision Transformer~\cite{caron2021emerging}. 
Our results demonstrate the effectiveness of all three \MyMethod variants. Specifically, we assess \MyMethod in a real applicability authorization scenario, preventing a pre-trained encoder from being misused for military purposes while keeping its generalizability to civilian ones.
\end{enumerate}

\section{Background} \label{sec: related works}
\subsection{Pre-trained Encoders and Model Probing} \label{sec: representation learning}
Pre-trained models are widely used in computer vision~\cite{marcelino2018transfer, parisi2022unsurprising, yuan2021florence}, representation learning~\cite{bengio2013representation, shen2021financial, yi2022graph, zhong2016overview}, and natural language processing~\cite{han2021pre, qiu2020pre, wang2022pre}, which embed pre-learned knowledge as the model initialization to reduce the complexity in training new tasks.
Taking transfer learning in vision tasks as an example, there are three common strategies:
\yf{why vision specifically here? can it be omitted?} \xxu{I share the same concern mentioning vision in a specific way... Can we say there are three strategies, taking the vision tasks as an example?}
\rd{revised. I also revise the following method from visual prompting to prompting.}

\noindent\textbf{Full Fine-tuning}: Full fine-tuning leverages the entire pre-trained model as the training initialization and fine-tunes it with the target dataset.  
It often has good performance but has the risk of stability and catastrophic forgetting~\cite{chen2019catastrophic, tajbakhsh2016convolutional}.\\
\noindent\textbf{Prompting}: Rather than finetune the model parameters, prompting redirects the pre-trained model via modification on the inputs (i.e., visual prompt). Prompting is efficient but performance experiences a larger degradation~\cite{bahng2022exploring, gan2023decorate, jia2022visual, yang2024fine}.
\noindent\textbf{Model Probing}:
Probing freezes the early layers of the pre-trained model (e.g., deep convolutional layers or self-attention layers~\cite{cordonnier2019relationship, han2022survey}) as the fixed pre-trained encoder and fine-tunes the downstream classifier. It has a small training cost and high stability of the training process~\cite{belinkov2022probing, gao2023tuning, ravichander2020probing, wu2024structured}.

\yf{you need a summary line here to state why you focus on model probing, instead of other two methods} \rd{added}

In this work, we focus on model probing as it is more efficient and stable than fully fine-tuning and has better performance than prompting.
Probing also supports pre-trained encoders from different training schemes, which can be categorized into three types: \textit{supervised}, \textit{unsupervised}, and \textit{self-supervised}.
For supervised learning, the model (i.e., encoder and downstream head) is trained directly using labeled training data and a loss function (e.g., cross-entropy loss)~\cite{cunningham2008supervised}.
Unsupervised learning aims to learn from unlabeled data, using methods such as Gaussian Mixtures Model (GMM)~\cite{reynolds2009gaussian}, Variational Autoencoder (VAE)~\cite{kingma2013auto}, and Generative Adversarial Network (GAN)~\cite{goodfellow2020generative}.
Self-supervised learning aims to train an encoder to predict one part of data given another part of the input~\cite{chen2020simple,  he2020momentum, jaiswal2020survey}. 
It leverages the inherent data characteristics and shows increasing robustness and generalizability of the encoder~\cite{hendrycks2019using}. 
Specifically, given the input (e.g., an image), one will use data augmentation operations (e.g., cropping, color jitter, and adding random noise) to build augmented images. 
The training objective is to make the encoder generate similar embeddings for augmented images from the same input, denoted as positive pairs; while ensuring the discrepancy of embeddings from different images, denoted as negative pairs.
The training of a self-supervised encoder utilizes contrastive loss~\cite{chen2020simple, he2020momentum}, which increases the similarity between positive pairs but decreases those of negative pairs.
Our design of \MyMethod considers various data accessibility of prohibited domains, which aligns with the training process of pre-trained encoders--with or without labeled datasets.


\subsection{Applicability Authorization} \label{sec: bg-ip}
Recently, applicability authorization, a new IP protection scheme, has been proposed to address the rising concerns of IP infringement on DNN models~\cite{wang2023model,wang2022non,zhou2024archlock, ding2025non}. 
Traditional model IP protection aims to protect the rights of owners of DNN models with two typical defense strategies: ownership verification and usage authorization. 
Ownership verification is designed to trace the illegal behavior of IP infringement using methods such as embedding watermarks during the training procedure or recording fingerprints of the model owner~\cite{jia2021entangled,yang2021robust,zhang2018protecting}. 
In contrast, usage authorization aims to restrict user access to the model, ensuring that only verified, trusted users can access with assigned authorization keys~\cite{alam2020deep, chakraborty2020hardware}.
Instead of protecting the model parameters or hyper-parameters directly like traditional methods, applicability authorization focuses on the unauthorized transfer of the pre-trained models \cite{zhou2024archlock}.
Specifically, it aims to prevent malicious transfer learning through which an attacker can abuse the pre-trained model for prohibited data or tasks, i.e., non-transfer-learning.
In this work, we further propose \MyMethod to address the challenges of applicability authorization of pre-trained \textit{encoders} to safeguard them from unauthorized probing.

\subsection{Non-Transferable Learning (NTL)} \label{bg-ntl}
Wang et al.~\cite{wang2022non} introduced NTL for applicability authorization of an entire model without any fine-tuning.
In particular, NTL leverages a negative regularization term on the model's target domain performance:
{\small
\begin{equation}
    L_{NTL} = L_\mathcal{S} + R_\mathcal{T} \label{eq: ntl objective}
\end{equation}
}where $L_\mathcal{S}$ is the Kullback–Leibler (KL) divergence/loss on the source dataset, aiming to retain the model's performance on the source domain. Model non-transferability comes from the regularization term, defined as $R_\mathcal{T}=- \text{min}(\beta, \alpha \cdot L_\mathcal{T} \cdot L_{dis})$~\cite{wang2022non}, where $L_\mathcal{T}$ is the KL loss on the target dataset, $L_{dis}$ measures the feature space distance between the source and target domains (using Maximum Mean Discrepancy), and $\alpha$ and $\beta$ are scaling factors. Another prior work~\cite{wang2023model} proposes an additional CUTI-domain for regularization on private style features with the $R_\mathcal{T}$ as $-L_\mathcal{T}$.
In addition, previous works also proposed `source-only' NTL for cases when there is no target data available. 
As the term `source-only' indicates, this strategy leverages generative models (i.e., GAN) to create a synthetic dataset, which serves as the boundary from the  source domain to prohibit the model's transferability to all other domains.

\subsection{Limitations of Prior Works} \label{bg-prior}
Prior works focus on the case when the attacker uses the trained model directly but cannot fine-tune it~\cite{wang2023model, wang2022non}. 
However, with the increasing popularity and low cost of probing the pre-trained (fixed) encoder, the applicability authorization (model non-transferability) can be bypassed in a few probing epochs.
Moreover, previous methods predominantly add a regularization term solely based on the model outputs. 
Although \cite{wang2022non} introduces a feature space distance as a regularization between the source and target domains, they cannot ensure restriction as the class discrepancy might still be large on the feature space.
Furthermore, previous methods only consider the case of supervised NTL, i.e., the defender has access to one labeled target dataset and one labeled source dataset. 
Our Challenge 3 is closer to the practical scenario where the defender lacks knowledge about the prohibited target.
Pre-trained encoders and model probing bring new challenges in data availability for applicability authorization. 
Our work \MyMethod aims to address them accordingly.

\section{Threat Model} \label{sec: threat model}
In this work, we tackle applicability authorization for pre-trained encoders, aiming to prevent malicious users from probing the encoder for harmful tasks (i.e., unethical, illegal, or abusive activities). 
In this paper, we focus on vision encoders and image classification as the downstream task.

\subsection{Malicious Users} \label{sec: malicious user}
The attackers are users with malicious intent to breach the usage policy of fixed pre-trained encoders with probing~\cite{gu2020self}.

\noindent\textbf{Objective.} Their objective is to exploit pre-trained encoders for tasks that are not allowed, specifically,
accurately classifying samples from prohibited domains.
Other forms of DNN IP infringement of the pre-trained encoder, e.g., model stealing attacks, are out of the scope of this paper.  

\noindent\textbf{Capabilities.} Capabilities of the malicious users include:
\begin{itemize}[leftmargin=*]
    \item They can probe a pre-trained encoder using inputs from prohibited target domains and utilize the representations to train a local downstream classifier for inference. 
    Although they query the encoder (as a service or local private model), it is a black-box with both structure and parameters unknown. 
    \item Users can build their own downstream classifiers, customizing the classifier's hyper-parameters and fine-tuning the parameters with any learning rates and optimizers. 
    \item Following the common setting of probing, we assume that the attacker has a small amount of data from the prohibited domain for fine-tuning (e.g., $10\%$ from the target domain).
\end{itemize}

\begin{figure*}[t]
    \centering
    \includegraphics[width=\linewidth]{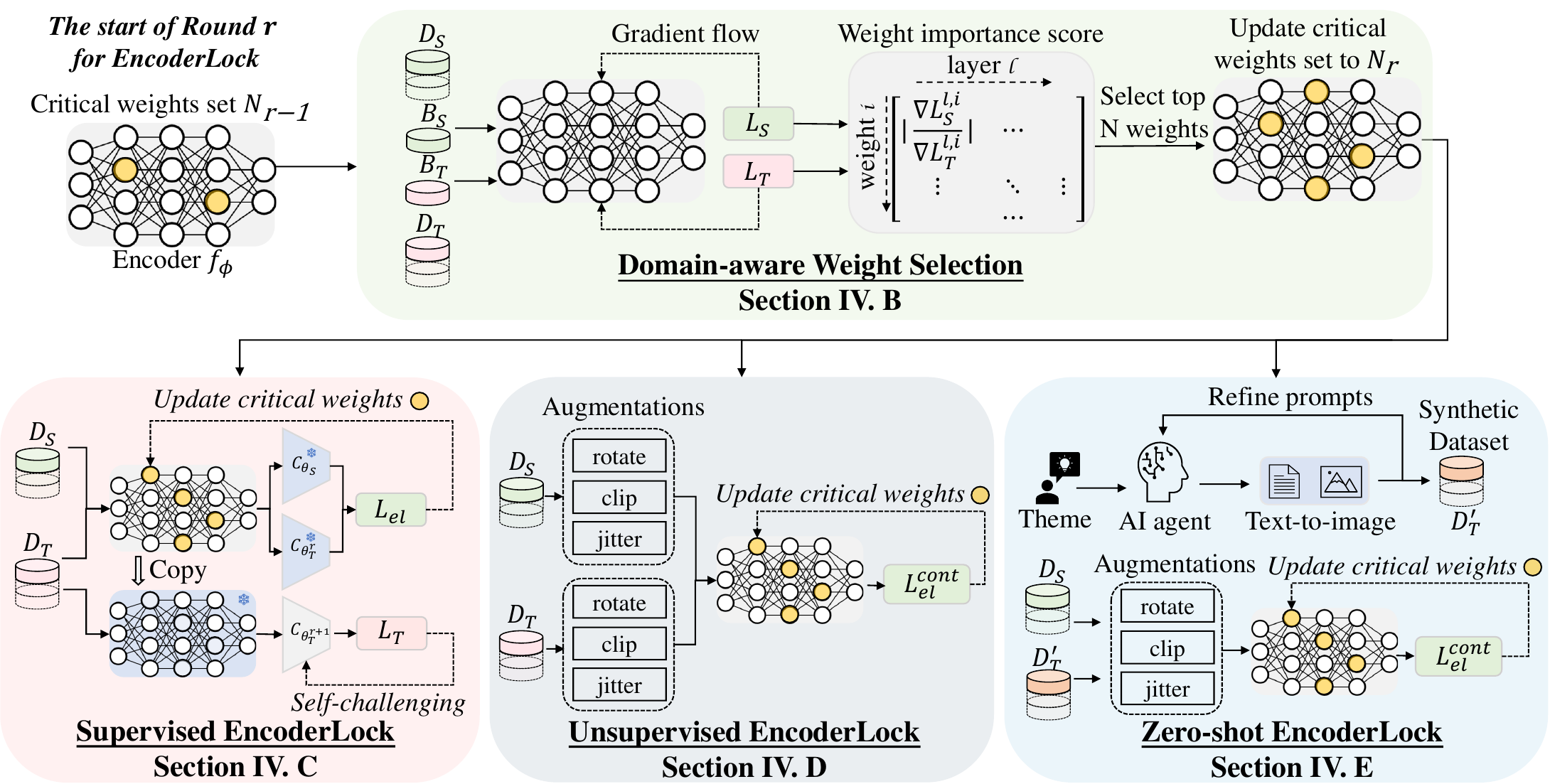}
    \caption{\textbf{Overview of the  proposed \MyMethod framework and paper organization}. The procedure in Round $r$ includes: 1.\textit{domain-aware critical weight selection algorithm}: take data batches $\mathcal{B}_\mathcal{S}$ and $\mathcal{B}_\mathcal{T}$ from the authorized source dataset $\mathcal{D}_\mathcal{S}$ and the prohibited target dataset $\mathcal{D}_\mathcal{T}$, respectively, and calculate the weight importance with gradients of loss $L_\mathcal{S}$ and $L_\mathcal{T}$ and choose critical weights to update for the round $r$ as $N_r$, note here specific losses depend on different levels of accessibility of the target domain;
    2. \textit{\MyMethod weight update algorithm} (with three variants for the three levels of target domain dataset), utilizing the supervised \MyMethod loss $L_\text{el}$, unsupervised contrastive loss $L_\text{el}^{\text{cont}}$ and the generated synthetic dataset $D_\mathcal{T}'$, respectively.}
    \yf{where is self-challenging learning?  the framework misses the outer loop?}
    \rd{This figure is for one loop, I try to add the loop in previous version, the figure becomes not very clean. I think draw one loop is fine here as we mention it in the caption. The self-challenging is only for the supervised encoderlock (because we need the labeled dataset to train the target classifier), I will add it.}
    \label{fig: encoderlock framework}
\end{figure*}

\subsection{Model Owner} \label{sec: eaas owners}
Model owners aim to safeguard the pre-trained encoder against malicious probing proactively. 
Following the common definition of transfer learning, the dataset on which the encoder is trained is defined as the \textbf{source domain (authorized domain)}, and the dataset that the (malicious) downstream classifier is trained for is the \textbf{target domain (prohibited domain)}.
Moreover, we define data domains other than these two as \textbf{admissible domains}, indicating that the usage of the encoder on these domains is allowed but their performance is not guaranteed like the source domain. 

\noindent\textbf{Protection Objective.} 
The major goal includes: restricting the pre-trained encoder from being probed for the prohibited domain; preserving its performance on the authorized domain.

\noindent\textbf{Capabilities.} The capabilities of the model owner include:
\begin{itemize}[leftmargin=*]
    \item The model owner has full control of the encoder to adjust the architecture, hyper-parameters, and parameters, and manage training strategies for the encoder before deploying it. 
    \item The owner has no access to the user's dataset after deployment to detect if the probing samples belong to the prohibited domain, and has no knowledge about the probing process or downstream heads. 
    \item The owner may have different levels of accessibility to the prohibited domain, from high to low:
     a) \colorbox{koblue!20}{Level 1}: The owner has a labeled target dataset; 
     b) \colorbox{darkblue!20}{Level 2}: The owner obtains the target dataset, but it is unlabeled;
    c)  \colorbox{kopink!20}{Level 3}: The owner only has an abstract concept about the prohibited target domain (i.e., a text description), which is called a `theme'.
    We propose three variants of EncoderLock to address different levels of accessibility, respectively.
\end{itemize}
\section{Proposed Framework: \MyMethod} \label{sec: method}
In this work, we propose a new applicability authorization strategy for a pre-trained encoder against malicious probing, which we call {\em \MyMethod}. 
Fig.~\ref{fig: encoderlock framework} depicts an overview of the framework, which consists of two major steps - domain-aware weight selection algorithm and specific weight updating algorithms catering to the different levels of accessibility of the target domain. 
By managing the weights, \MyMethod restricts the encoder from being probed on the prohibited target domain to extract useful information, while ensuring that the encoder correctly responds to authorized (source) inputs. 
Existing literature mostly focuses on protecting pre-trained models from being transferred~\cite{wang2023model, wang2022non}, while our \MyMethod targets pre-trained encoders, with several unprecedented challenges outlined in Section \ref{sec: design objective}.

\subsection{Design Objectives and Challenges for \MyMethod} \label{sec: design objective}
In addition to the traditional design objective of controlling the transferability of pre-trained models to prohibited target domains~\cite{wang2022non}, \MyMethod faces three additional challenges that need to be effectively addressed.
\yf{why does EncoderLock inherit the traditional design objective? you don't want to make your work a superset of everything} \rd{I want to express that the encoderlock also need preserve the source performance and reduce the target. But under a different setting. Maybe we do not need to mention the traditional objective?}

\noindent\textbf{Challenge 1. Preservation of Integrity:} 
The integrity of the encoder lies in maintaining the pre-learned knowledge about the authorized domains.  One question is raised:  how can \MyMethod make minimal modifications to the encoder to restrict it on the prohibited domain while preserving the integrity on the source domain?

\noindent\textbf{Challenge 2. Robustness to malicious probing:} 
When malicious users adjust the downstream heads for the prohibited domain with any learning rate and optimizer, how to successfully `lock' the encoder against malicious probing?

\noindent\textbf{Challenge 3. Different target domain data accessibility:} 
In reality, as the defender (model owner) may have various levels of knowledge about the target domain, how should \MyMethod be designed for practical scenarios including unlabeled datasets or even no samples from prohibited domains?

\subsection{Domain-aware Weight Selection} \label{sec: critical weight search} 
\begin{figure}[t]
    \centering
    \includegraphics[width=0.7\linewidth]{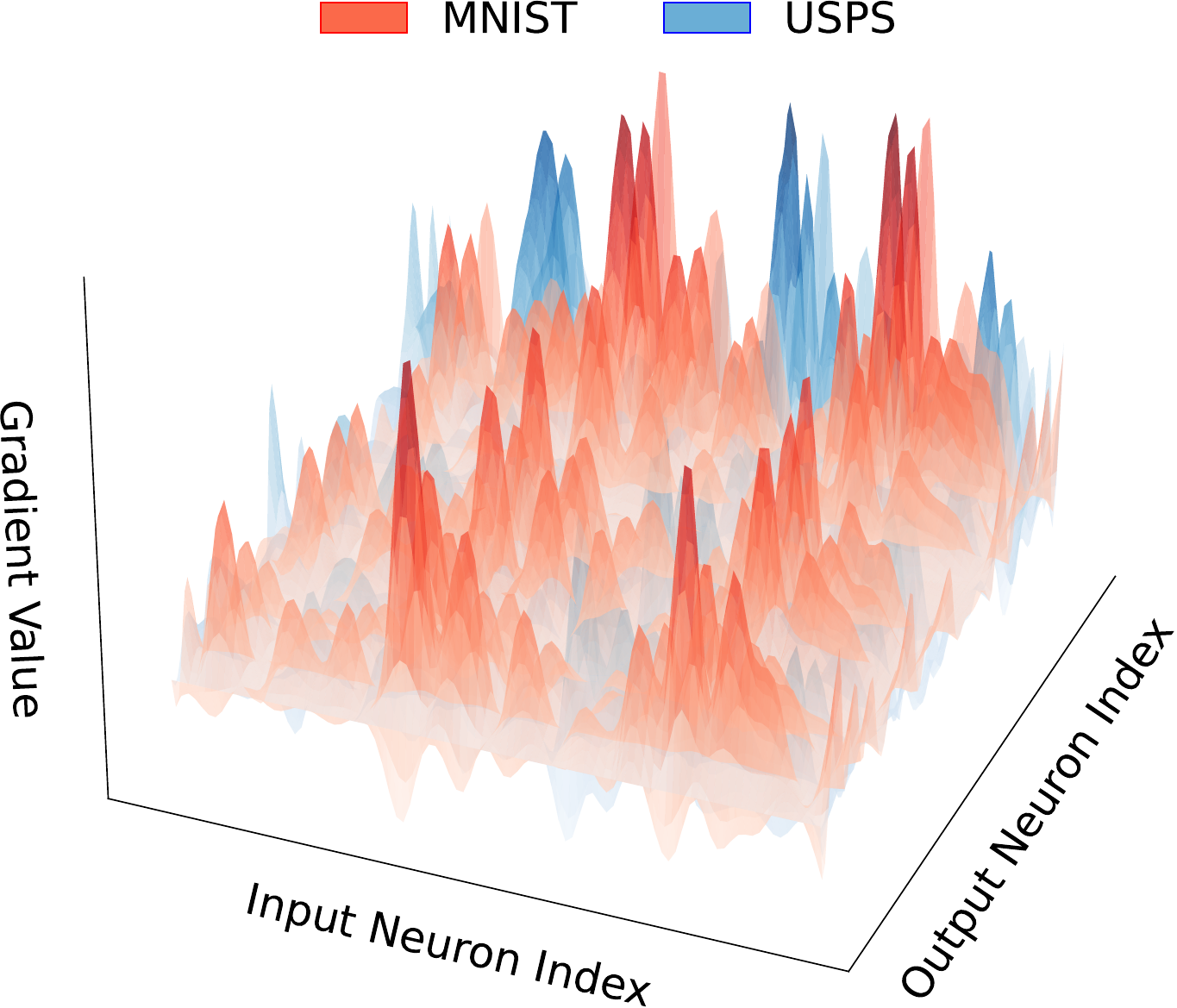}
    \caption{\textbf{Visualization of weight importance in a pre-trained model---}The X-Y plane represents the weight matrix of a selected dense layer in a model trained on MNIST and probed for USPS. 
    The color and height indicate each weight's importance to the output (the higher and darker, the more important).
}
    \label{fig: motivation-dfa}
\end{figure}

To address \textbf{Challenge 1}, we propose a domain-aware weight selection strategy to selectively update weights that are critical only for the target domain, thereby minimizing \MyMethod's negative impact on the integrity of the pre-trained encoder.
Our strategy is motivated by two observations of DNN models: 
1) \textit{Weight importance varies across different domains}.
Different sets of critical weights in the same model may respond to different domains, which can be measured by the weight's gradient magnitude.  
This notion has been exploited in achieving domain-specific pruning~\cite{liu2021transtailor}, effective fault injections on DNN parameters~\cite{liu2017fault, rakin2019bit, zhao2019fault}, and watermarking embedding~\cite{yang2021robust}.
Fig.~\ref{fig: motivation-dfa} demonstrates one example of different weight importance for different domains. The model is a 
multi-layer perceptron network trained on MNIST (source domain),
and the distribution of the weight gradients is shown in red color. When this model runs inference for another dataset USPS (target domain), the weights gradient profile is shown in blue color, distinctly different from that on MNIST. 
2) \textit{Over-parametrization}---DNNs often have more weights than required, with a large portion being insignificant. 
This characteristic is widely utilized in model compression for efficiency~\cite{frankle2018lottery, liu2018rethinking, zhu2017prune}, where only critical weights are retained while others are pruned away. 

Our {\em Domain-aware Weight Selection} (DWS) algorithm is  described in Function \ref{alg: critical_weight_search}. 
The search process runs iteratively.  
In the $r^{\text{th}}$ round, it uses datasets from the source and target domains, $\mathcal{D}_\mathcal{S}$ and $\mathcal{D}_\mathcal{T}$, to search for and update the critical weight set $\mathcal{N}_r$.
It is important to note that the composition of $\mathcal{D}_\mathcal{T}$ depends on the level of data accessibility. 
For supervised \MyMethod (Level-1), inputs from the target dataset have labels, and the loss $L_\mathcal{T}$ is calculated using cross-entropy, similar to $L_\mathcal{S}$. In the unsupervised and zero-shot \MyMethod (Level-2 and Level-3), the target dataset consists of  unlabeled or synthetic images, and we propose using a contrastive loss for these unlabeled inputs, presented in Eq.\eqref{eq: unsupervised loss}.
The weight importance score is defined as the magnitude ratio of gradients for the $i^{th}$ weight in layer $l$ between the target and source domains $|\nabla L^{l,i}_{\mathcal{T}}/\nabla L^{l,i}_{\mathcal{S}}|$. 
This score is used to identify weights that are critical to target domains but less crucial to the source. 

The search process is iterated across \( R \) rounds, with \( N \) weights selected in each round, resulting in a total of \( N \times R \) weights to update. The values of \( N \) and \( R \) are two hyperparameters that control the number of altered weights and will be discussed further in Section~\ref{sec: ablation}.
Such design allows us to process the datasets in batches and implement the self-challenging training scheme (see Section \ref{sec: self-challenging training scheme}).

In Section~\ref{sec: supervised encoderlock} to \ref{sec: method: zero-shot encoderlock}, we further discuss more details about how to update the selected weights and how to achieve robustness against downstream fine-tuning (\textbf{Challenge 2}).
The method varies across different levels of accessibility to the target domain, as shown in three branches of Fig.~\ref{fig: encoderlock framework}.

\floatname{algorithm}{Function}
\begin{algorithm}[t]
\caption{Domain-aware Weight Selection 
}\label{alg: critical_weight_search}
\begin{algorithmic}[1]
    \Require Source domain $\mathcal{D}_{\mathcal{S}}$, Target domain $\mathcal{D}_{\mathcal{T}}$, Pre-trained encoder parameters $\phi$, Number of new critical weights $N$, Set of critical weights for the previous round $\mathcal{N}_{r-1}$
    \Ensure Set of critical weights $\mathcal{N}_r$ 
    \Function{DWS}{$\mathcal{D}_{\mathcal{S}}$, $\mathcal{D}_{\mathcal{T}}$, $\phi$, $\mathcal{N}_{r-1}$, $N$}
        \State Sample a training batch $\mathcal{B}_{\mathcal{T}}$ from $\mathcal{D}_\mathcal{T}$
        \State Sample a training batch $\mathcal{B}_{\mathcal{S}}$ from $\mathcal{D}_\mathcal{S}$
        \Statex \textcolor{kopink}{\quad /* Compute gradients for both batches */}
        \State $\nabla L_{\mathcal{T}} \gets$ ComputeGradients($\phi$, $\mathcal{B}_{\mathcal{T}}$)
        \State $\nabla L_{\mathcal{S}} \gets$ ComputeGradients($\phi$, $\mathcal{B}_{\mathcal{S}}$)
        \Statex \textcolor{kopink}{\quad /* Compute scores and select critical weights */}
         \State Compute score for $i^{th}$ weight in $l^{th}$ layer:   $\left |\frac{\nabla L^{l,i}_{\mathcal{T}}}{\nabla L^{l,i}_{\mathcal{S}}}\right|$
        \State Select top $N$ weights: $\argmax_N\left|\frac{\nabla L^{l,i}_{\mathcal{T}}}{\nabla L^{l,i}_{\mathcal{S}}}\right| $ 
        \State $\mathcal{N}_r \gets \mathcal{N}_{r-1} \cup \{\text{selected weights}\}$
        \State \textbf{return} $\mathcal{N}_r$
    \EndFunction
\end{algorithmic} 
\end{algorithm}

\subsection{Supervised \MyMethod} \label{sec: supervised encoderlock}
Level 1 \MyMethod is supervised, with a labeled target domain dataset. 
Specifically, given the source domain $\mathcal{D}_{\mathcal{S}}$ and target domain $\mathcal{D}_{\mathcal{T}}$, with $(x_{\mathcal{S}}, y_{\mathcal{S}}) \in \mathcal{D}_{\mathcal{S}}$ and $(x_{\mathcal{T}}, y_{\mathcal{T}}) \in \mathcal{D}_{\mathcal{T}}$ 
as the corresponding datasets, let $f_{\phi}$ denote the pre-trained encoder, and $C_{\theta_{\mathcal{S}}}$ and $C_{\theta_{\mathcal{T}}}$ denote the auxiliary downstream task classifiers for the source and target domains, respectively. Our objective is to find an optimal encoder $\phi^*$ 
that minimizes $L_{\mathcal{S}}$ but maximizes $L_{\mathcal{T}}$, which are expressed as:
{\small
\begin{equation}
L_{\mathcal{S}} = L(C_{\theta_{\mathcal{S}}}(f_{\phi}(x_{\mathcal{S}})), y_{\mathcal{S}})
\end{equation}
\begin{equation}
L_{\mathcal{T}} = L(C_{\theta_{\mathcal{T}}}(f_{\phi}(x_{\mathcal{T}})), y_{\mathcal{T}})
\end{equation}
}
where $L$ is the classification loss function (i.e., cross-entropy loss) and is used to compute gradient in Algorithm~\ref{alg: critical_weight_search} for weight selection.
To restrict the impact on the encoder's generalizability, we require $\|\phi^*-\phi\|_0 \le M$ (:=$N \times R$), where $\|  \cdot \|_0$ is $\ell_0$ norm and $M$ signifies the weight change budget. 

The fundamental supervised \MyMethod consists of three steps: 1) Domain-aware weight selection, 2) Non-transferability updating, and 3) Self-challenging downstream model training. 
We run these three steps iteratively for \(R\) rounds or until the accuracy of the auxiliary downstream classifier reaches the early stopping criterion.
For the three different levels of target domain data accessibility, the weight search and update algorithms are similar, but with different loss functions. 
But the supervised \MyMethod, with its loss design for the output space, requires an additional self-challenging training step to ensure its robustness.
We next discuss the other two design steps for supervised \MyMethod in detail.

\subsubsection{Weight Updating for Non-transferability} 
With critical weights selected to update, we design a loss function in the form of Equation \eqref{eq: ntl objective}, 
focusing on the regularization term $R_\mathcal{T}$ to mitigate the malicious probing for the target domain.
Previous regularization terms \cite{wang2023model, wang2022non} only consider the target domain, which leads to unstable performance especially when $L_\mathcal{S}$ and $L_\mathcal{T}$ are at different orders of magnitude.
In particular, when $L_\mathcal{S}$ is very small (i.e., near zero), the introduction of $R_\mathcal{T}$ will cause a strong impact on $L_\mathcal{S}$.
Therefore, we propose a new log-ratio regularization term considering both $L_\mathcal{S}$ and $L_\mathcal{T}$:
{\small
\begin{equation}
    L_{el}=L_\mathcal{S} + R_\mathcal{T}, \: \textnormal{where} \: R_\mathcal{T}=\log(1 + \alpha \frac{L_{\mathcal{S}}}{L_{\mathcal{T}}})
\label{eq: regularization}
\end{equation}
}
Such logarithmic regularization term gently penalizes the loss ratio between the source and target, with \( \alpha \) moderating the balance between preserving the source domain accuracy and enforcing the target domain non-transferability.
Consequently, the optimization objective for the encoder is defined as:
{\small
\begin{equation}
    \phi^*= \arg\min_{\phi} 
    L_{el}(\phi,\theta_\mathcal{S},\theta_\mathcal{T}) \quad
 \text {  s.t.}  \enspace \|\phi^*-\phi\|_0 \le M \enspace {\forall \theta_\mathcal{S},\theta_\mathcal{T}}
\label{eq: loss}
\end{equation}
}

\subsubsection{Self-challenging Training Scheme} \label{sec: self-challenging training scheme}

\floatname{algorithm}{Algorithm}
\setcounter{algorithm}{0}
\begin{algorithm}[t]
\caption{Self-challenging Training Scheme}\label{alg: target_domain_counter_model}
\begin{algorithmic}[1]
    \Require Pre-trained encoder with $\phi$, Source domain $\mathcal{D}_{\mathcal{S}}$,  Target training dataset $\mathcal{D}^{\textbf{train}}_{\mathcal{T}}$, Target validation dataset $\mathcal{D}^{\textbf{valid}}_{\mathcal{T}}$, Number of critical weights $N$, Number of rounds $R$, Desired target accuracy $\alpha_{goal}$.
    \Ensure Encoder with supervised \MyMethod  $\phi^*$
    \For{$r = 1$ to $R$}
        \Statex \textcolor{kopink}{\quad /* Initialize critical weights set for the first round*/}
        \If{$r==1$}
        \State Initialize set $\mathcal{N}_{r-1} \gets \emptyset$
        \EndIf
        \Statex \textcolor{kopink}{\quad /* Begin Domain-aware Weight Selection */}
        \State $\mathcal{N}_r$ = DWS($\mathcal{D}_{\mathcal{S}}$, $\mathcal{D}_{\mathcal{T}}$, $\phi$, $\mathcal{N}_{r-1}$, $N$)     
        \Statex \textcolor{kopink}{\quad/* Minimax optimization for enhancing robustness */}
        \State $\phi^{*} \gets$ optimize weights in $\mathcal{N}_{r}$ to minimize  $ L_{el}$ \eqref{eq: loss}\yf{here is it (6) or (5) or (4)?} \rd{(4) is the formula for el, (5) is the formula for updating the weights. Maybe (5) is better here?}
        \State Initialize an auxiliary downstream $C_{\mathcal{T}}(\cdot;\theta_{\mathcal{T}})$        
        \State Fine-tune $C_{\mathcal{T}}$ using $\mathcal{D}^{\textbf{train}}_{\mathcal{T}}$ with encoder $\phi^{*}$ 
        \State Compute accuracy $\alpha_{\mathcal{T}}$ of $C_{\mathcal{T}}$ on $\mathcal{D}^{\textbf{valid}}_{\mathcal{T}}$
        \Statex \textcolor{kopink}{\quad /* Stop Criterion */}
        \If{$\alpha_{\mathcal{T}} < \alpha_{goal}$ or $\|\phi^*-\phi\|_0 > N \times R$} 
    \State \textbf{return} $\phi^*$
    \EndIf
    \EndFor
\end{algorithmic}  
\end{algorithm}

To update the encoder weights in supervised \MyMethod following Equation 
\eqref{eq: loss}, we consider the auxiliary downstream classifier to compute $L_\mathcal{T}$. However, malicious users have full control of the downstream classifier, including adjusting the architecture and choosing the optimization method, and can fine-tune the model parameters based on the extracted features. Consequently, 
the performance of the pre-trained encoder and the classifier on the target domain can be improved. 
Relying solely on a fixed-weight target classifier could lead to vulnerability, where supervised \MyMethod may only be non-transferable for given auxiliary classifiers but not for others the malicious user opts for.

To improve robustness against any potential malicious probing for supervised \MyMethod, we propose a self-challenging training scheme with  
a minimax problem formulation as:
{\small
\begin{equation}
    \phi^*= \arg\min_{\phi}\max_{\theta_\mathcal{T}} 
    L_{el}(\phi,\theta_\mathcal{S},\theta_\mathcal{T}) \quad
 \text { s.t. }  \enspace \|\phi^*-\phi\|_0 \le M
\label{eq: min max loss }
\end{equation}}Specifically, during the iterations of updating critical weights in the encoder part, we also adjust the target downstream models iteratively. 
It is noted that the training objective of the target downstream models will be adversarial to \MyMethod's applicability objective---the fine-tuning aims to extract useful features in the embeddings to enhance the target domain performance. 
The retrained downstream model adjusts itself frequently to create a \textit{challenging} target downstream classifier that will increase $L_{el}$ (decreasing $L_\mathcal{T}$), prompting the supervised \MyMethod to adjust more critical weights on the encoder part. 
Algorithm~\ref{alg: target_domain_counter_model} outlines this self-challenging training process. 
To ensure the randomness of the target downstream classifiers, every iteration we retrain it from scratch (with a random initialization). 
The iterative training proceeds until the target downstream classifier's accuracy drops below a predefined threshold or reaches the maximum number of altered weights $M$. 
The self-challenging training scheme ensures a gradual and smooth reduction in the encoder's transferability, forcing the encoder part to extract features that are less useful for the target domain, thereby leading to more robust performance even when the attacker probes the downstream model adaptively.

\subsection{Unsupervised \MyMethod} \label{sec: method: unsupervised encoderlock}
\begin{figure}[t]
    \centering
    \includegraphics[width=0.95\linewidth]{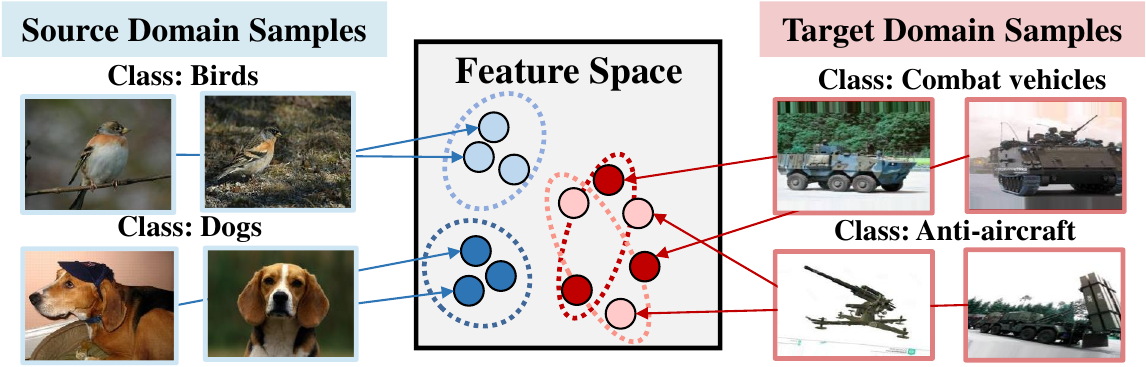}
    \caption{\textbf{Design motivation of unsupervised \MyMethod}}
    \label{fig: unsupervised-framework}
\end{figure}

In this section, we address Level-2 accessibility of the target domain via \textit{unsupervised \MyMethod}.
This scenario is practically relevant when the goal is to prevent transferring to arbitrary sets of images while getting their labels is either infeasible or expensive.  
Our method leverages the technique from self-supervised representation learning~\cite{chen2020simple, he2020momentum}, which 
builds a highly distinguishable feature space without labeling.

The design idea for unsupervised \MyMethod is as follows: for the latent embeddings of samples from the source domain, we aim to ensure their high discrepancy between classes; while for those from the prohibited target domain, our objective is to obfuscate the latent clusters boundary so that the embeddings would not contain much information about the class. As shown in Fig.~\ref{fig: unsupervised-framework}, an expected encoder will automatically cluster the samples from the source domain but blur the class boundaries of the target domain. Due to such direct manipulation towards the encoder's feature space, the unsupervised \MyMethod is always robust to different downstream heads and doesn't require further self-challenging training.

Towards this goal, we introduce a self-supervised regularization term $R_\mathcal{T}$ to be used in Equation \eqref{eq: regularization}. 
Specifically, given a batch of samples from the target domain, we leverage data augmentation, including random crop, color jitter, or Gaussian blur~\cite{shorten2019survey, mikolajczyk2018data}, to create a set of positive pairs and a set of negative pairs. 
Any pair with a sample and an augmented sample from the same original image is defined as positive, and we denote their feature space as $(z_i, \tilde{z}_i)$, where $z_i$ is defined as the normalized embedding of the sample $x_i$ using the encoder $f$. 
Any pair with augmented samples from different original images is defined as negative, denoted as $(z_i, \tilde{z}_j)_{i\neq j}$.\yf{should it be $\tilde{z}_i$ here?} \rd{It should be $\tilde{z}_j$, it is a pair current latent sample $z_i$ with the augmented sample originally different from $z_i$, corresponding to the positive pairs.} \yf{I mean the first one, not the second one, with tilde indicating it is from an augmented image?}\rd{Yes, tilde means the embedding comes from an augmented input}\yf{should it be $(\tilde{z}_i, \tilde{z}_j)_{i\neq j}$ here?}\rd{I see. They have similar performance. As we already minimize the distance between zi and tilde zi.}
We define the contrastive loss function $L^{\text{cont}}$ as:
{\small
\begin{equation}
L^{\text{cont}} := -\frac{1}{N_B}\sum_{i=1}^{N_B} \text{log} (\frac{\text{sim}(z_i, \tilde{z}_i)}{\sum_{j=1}^{N_B} \text{sim}(z_i, \tilde{z}_j)})
\label{eq: unsupervised loss}
\end{equation}
}where $N_B$ is the batch size, and $\text{sim}(\cdot, \cdot)$ computes the cosine similarity between the normalized embeddings. We select pairs from $\mathcal{S}$ to compute $L^{\text{cont}}_{\mathcal{S}}$, and from $\mathcal{T}$ to compute $L^{\text{cont}}_{\mathcal{T}}$. They are used to compute gradients in Algorithm\ref{alg: critical_weight_search}.
\yf{why do you need to compute $L^{\text{cont}}_{\mathcal{S}}$ in the same way as $L^{\text{cont}}_{\mathcal{T}}$?} \rd{We will use $L^{cont}$ the same way as the formula (4) for supervised's cross entropy loss.}\yf{I mean you can calculate the $L_S$ with formula (2), instead of formula (7), because the source dataset is always labeled.} \rd{I didn't consider that. It should also work. }

The presented loss function aims to increase the similarity between any positive pairs but reduce what between negative pairs, effectively pushing the encoder to learn representations that clearly distinguish similar samples from dissimilar ones within feature space. We follow the regularization framework in Eq.~\eqref{eq: regularization} and penalize ratios between contrastive losses: 
{\small
$$R_\mathcal{T}^{\text{cont}}=\log(1 + \alpha \frac{L^{\text{cont}}_{\mathcal{S}}}{L^{\text{cont}}_{\mathcal{T}}})$$ 
}

For the unsupervised \MyMethod, self-challenging training is not necessary because this loss function directly penalizes the discrepancy of the feature space for the target dataset. 
Therefore, as shown in Fig.~\ref{fig: encoderlock framework}, the procedure of unsupervised \MyMethod in one round includes: 1) Domain-aware weight selection with $L^{\text{cont}}$; 2) Update Encoder's Non-transferability with $R_\mathcal{T}^{\text{cont}}$, without retraining the \textit{challenging} classifier.

 \begin{figure}[t]
    \centering
    \includegraphics[width=\linewidth]{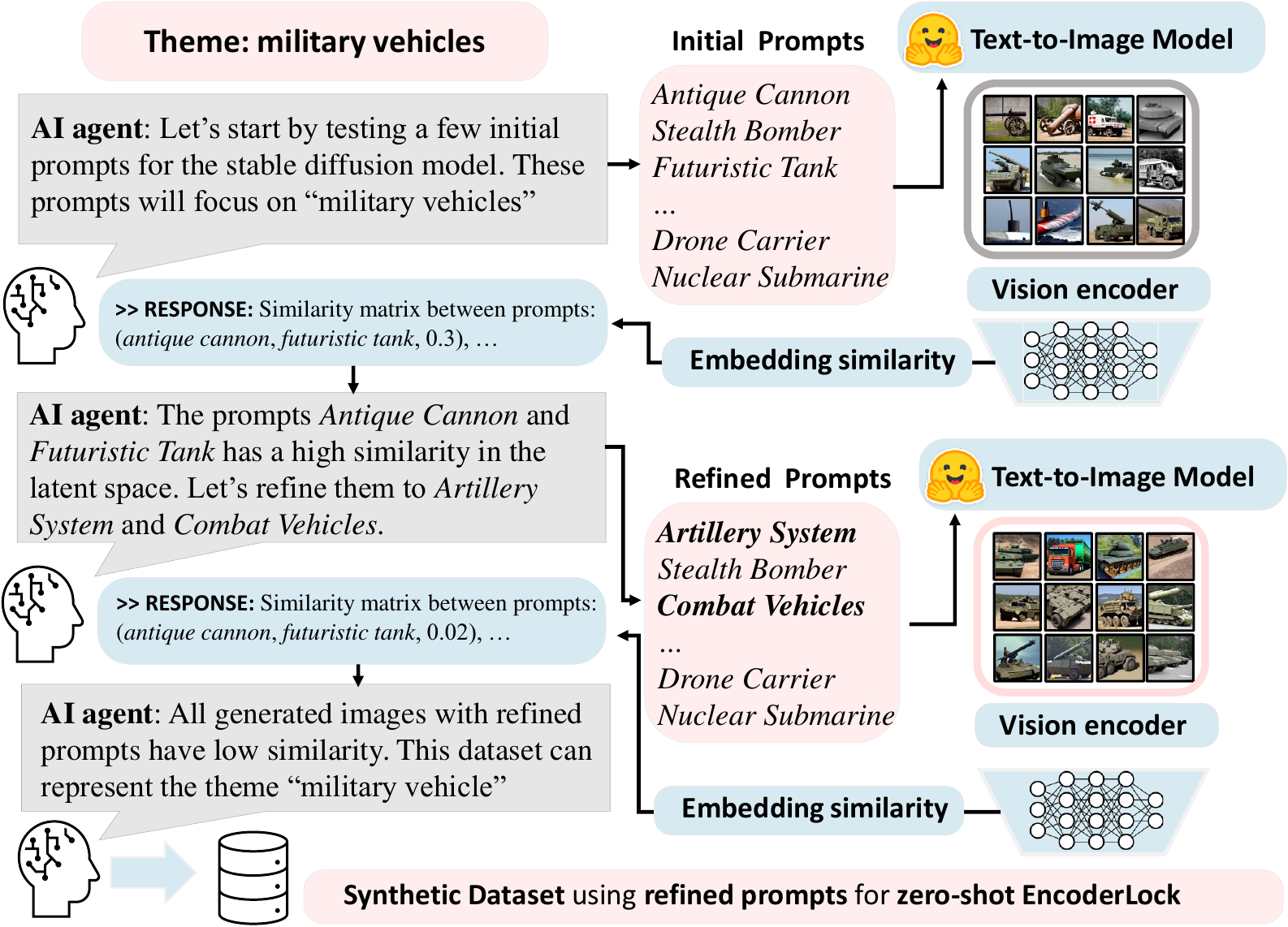}
    \caption{\textbf{Building synthetic datasets for zero-shot \MyMethod}}
    \label{fig: zero-shot-framework}
\end{figure}

\subsection{Zero-shot \MyMethod} \label{sec: method: zero-shot encoderlock}

In this section, we address Level-3 accessibility of the target domain for \MyMethod, where the model owner even has no target samples.  
This represents the most practical and relevant scenario, as the definition of harmful content is often vague in real-world applications.
For instance, in most cases, a DNN product's user guidelines regulate prohibited content using text descriptions of unethical or sensitive material.
How to turn such vague scope description into representative and comprehensive target domain dataset is a challenge. 
We define the basic knowledge about the target domain as  a \textit{theme}, which can be in the form of a text description, keywords, or reference figures. 
Using the target theme, zero-shot \MyMethod aims to generate a synthetic dataset for applicability authorization without relying on real-world samples or labels.
\yf{this sounds reasonable. but what is the major challenge? how to make sure the synthetic dataset is representative or comprehensive? what would be the requirement on the synthetic datasets. It is important for such arguments and reasoning before the specific techniques and methods of generating synthetic datasets.}
\rd{The way I try to add some requirements is the refinement process. We encourage a wide coverage of the synthetic dataset. Do you think that's enough? We also show that such dataset will have a degradation on both source and target in the experiment. That meets our design.}

Fig.~\ref{fig: zero-shot-framework} presents the framework of zero-shot \MyMethod, illustrating the process of generating a synthetic dataset for ``\textit{military vehicles}''. Section~\ref{exp: zero-shot encoderlock} will showcase the full results.
First, we employ a large language model (e.g., GPT-4~\cite{achiam2023gpt}) as an AI agent to generate text inputs, known as prompts, for the given theme. These prompts are then fed into pre-trained text-to-image models (e.g., CLIP~\cite{radford2021learning} and Stable Diffusion~\cite{rombach2021highresolution}) to generate the synthetic dataset.
To ensure the synthetic dataset comprehensively covers the target domain, we introduce a prompt refining framework. Using a pre-trained vision encoder, we extract latent features from the synthetic images and compute pairwise similarity scores between the initial prompts. This similarity matrix serves as feedback to the AI agent, enabling it to analyze the scores, identify similar prompts, and refine them accordingly.
For example, as shown in Fig.~\ref{fig: zero-shot-framework}, \textit{Antique Cannon} and \textit{Futuristic Tank} exhibit high similarity due to their shared barrel feature. Consequently, the AI agent revises these prompts to be \textit{Artillery System} and \textit{Combat Vehicles}. The refinement process continues until all prompt pairs demonstrate low similarity or the similarity stops decreasing.
Finally, we employ the synthetic dataset generated in the last round for unsupervised \MyMethod training, resulting in an encoder with restricted transferability to the target ``theme.''
\section{Experiments} \label{sec: experiments}
\subsection{Experiment Setup} \label{sec: experiment setup}
\begin{table}[t]
\centering
\caption{\textbf{Datasets used in evaluation of \MyMethod}}
\label{tab:dataset_summary}
\resizebox{\linewidth}{!}{
\begin{tabular}{l|l|l|c|c|c|c}
\hline
\textbf{Dataset} & \textbf{Abbr.} & \textbf{Type} & \textbf{Feat.} & \cellcolor{koblue!20}\textbf{Supervised} & \cellcolor{darkblue!20} \textbf{Unsupervised} & \cellcolor{kopink!20}\textbf{Zero-shot} \\
\hline
MNIST\cite{lecun1998gradient} & MT & digits & \multirow{ 7}{*}{\parbox{3cm}{Datasets that are used in the baselines\cite{wang2022non,wang2023model}. All samples are resized into $(32, 32, 3)$ and the label space is $10$.}} & $\surd$& $\surd$ & -  \\
\cline{1-3}
\cline{5-7}
USPS \cite{hull1994database} & UP & digits &  & $\surd$& $\surd$ & -  \\
\cline{1-3}
\cline{5-7}
SVHN\cite{netzer2011reading} & SN & digits &  & $\surd$& $\surd$ & -   \\
\cline{1-3}
\cline{5-7}
MNIST-M\cite{ganin2016domain} & MM & digits & &  $\surd$& $\surd$ & -   \\
\cline{1-3}
\cline{5-7}
Synthetic Digits\cite{ganin2015unsupervised} & SD & digits &  & $\surd$& $\surd$ & -   \\
\cline{1-3}
\cline{5-7}
CIFAR-10\cite{krizhevsky2009learning} & CF & image & & $\surd$& - & -    \\
\cline{1-3}
\cline{5-7}
STL-10\cite{coates2011analysis} & ST & image & & $\surd$& - & -   \\
\hline
EMNIST\cite{cohen2017emnist} & EM & char. & $47$-class characters & $\surd$& - & -   \\
\hline
CIFAR-100\cite{krizhevsky2009learning} & CF100 & image & $100$-class images & $\surd$& - & -   \\
\hline
ImageNette\cite{imagenette} & - & image & \multirow{3}{*}{\parbox{2.5cm}{High resolution images with the shape of $(224, 224, 3)$.}}   & $\surd$& $\surd$ & $\surd$ \\
\cline{1-3}
\cline{5-7}
ImageWoof\cite{imagenette} & - & image  &  & $\surd$& $\surd$ & $\surd$ \\
\cline{1-3}
\cline{5-7}
Military Vehicle\cite{bose_military_vehicles}\footnote{\url{https://www.kaggle.com/datasets/amanrajbose/millitary-vechiles}} & - & image &  & $\surd$& $\surd$ & $\surd$\\

\hline
\end{tabular}
}
\end{table}

\begin{table*}[t]
\centering
\setlength{\fboxsep}{1pt} 
    \caption{\textbf{\colorbox{koblue!20}{Supervised \MyMethod} performance: the encoder transferability---}pre and post-\MyMethod accuracy are reported, designated as `Before($\%$) $\Rightarrow$ After ($\%$)'. 
    Bold values show accuracies on the source domain
    }
    \resizebox{0.95\linewidth}{!}{
    \begin{tabular}{c|c|c|c|c|c|c|c|c}
    \hline
         \makecell{Source \textbackslash{} Target}& MT & UP & SN & MM & SD & $\Delta W$ & $Drop_\mathcal{S}$ & $Drop_\mathcal{T}$\\
     \hline
     & \multicolumn{8}{c}{\textbf{VGG-11}: 133M Parameters}  \\
     \hline
    MT & $\cellcolor{gray!20} \mathbf{99.53\Rightarrow 99.32}$ & $96.35\Rightarrow8.47$ & $43.74\Rightarrow18.98$ & $68.24\Rightarrow18.05$ & $69.65\Rightarrow13.67$ & 0.63\textperthousand  & $0.21\%\downarrow$ & $78.70\%\downarrow$ \\
    \hline
    UP & $97.70\Rightarrow11.35$ &\cellcolor{gray!20} $\mathbf{97.91\Rightarrow 94.94}$ & $58.23\Rightarrow16.86$ & $65.24\Rightarrow15.43$ & $87.10\Rightarrow16.93$ &  0.43\textperthousand  & $2.97\%\downarrow$ & $76.16\%\downarrow$ \\
    \hline
    SN  & $95.30\Rightarrow19.72$ & $92.68\Rightarrow15.89$ & \cellcolor{gray!20} $\mathbf{94.04\Rightarrow90.51}$ & $71.51\Rightarrow32.17$ & $96.96\Rightarrow 15.66$ & 2.50\textperthousand  & $3.53\%\downarrow$ & $76.62\%\downarrow$ \\
    \hline
    MM  & $98.85\Rightarrow12.71$ & $94.67\Rightarrow17.99$ & $53.80\Rightarrow23.80$ & \cellcolor{gray!20} $\mathbf{94.24\Rightarrow92.71}$ & $85.89\Rightarrow28.20$ & 0.99\textperthousand  & 1.53\%$\downarrow$ & $75.28\%\downarrow$ \\
    \hline
    SD& $97.13 \Rightarrow20.19$ & $93.57\Rightarrow18.68$ & $90.40\Rightarrow41.42$ & $71.53\Rightarrow 40.91$ &\cellcolor{gray!20} $\mathbf{99.83\Rightarrow98.89}$ & 1.78\textperthousand  & $0.94\%\downarrow$ & $64.13\%\downarrow$\\
    \hline
     & \multicolumn{8}{c}{\textbf{ResNet-18}: 11.4M  Parameters}  \\
    \hline
    MT & $\cellcolor{gray!20} \mathbf{99.48\Rightarrow99.12}$ & $93.77\Rightarrow13.16$ & $41.47\Rightarrow19.67$ & $70.02\Rightarrow20.77$ & $72.48\Rightarrow16.69$ & $0.31$ \textperthousand  & $0.28\%\downarrow$ & $71.73\%\downarrow$ \\
    \hline
    UP & $95.10\Rightarrow13.89$ & \cellcolor{gray!20} $\mathbf{96.11\Rightarrow95.69}$ & $33.72\Rightarrow11.36$ & $55.79\Rightarrow9.04$ & $61.76\Rightarrow16.53$ & $0.29$ \textperthousand  & $0.44\%\downarrow$ & $76.98\%\downarrow$ \\
    \hline
    SN  & $94.65\Rightarrow9.53$ & $88.04\Rightarrow15.65$ & \cellcolor{gray!20} $\mathbf{91.06\Rightarrow90.12}$ & $66.52\Rightarrow11.36$ & $95.08\Rightarrow10.45$ & $0.41$ \textperthousand  & $1.03\%\downarrow$ & $86.02\%\downarrow$ \\
    \hline
    MM  & $98.82\Rightarrow24.05$ & $92.33\Rightarrow12.81$ & $48.18\Rightarrow7.01$ & \cellcolor{gray!20} $\mathbf{91.49\Rightarrow90.39}$ & $78.03\Rightarrow18.53$ & $0.13$ \textperthousand  & $1.20\%\downarrow$ & $80.87\%\downarrow$ \\
    \hline
    SD & $96.76\Rightarrow10.39$ & $91.68\Rightarrow8.47$ & $86.74\Rightarrow30.70$ & $68.56\Rightarrow9.95$ & \cellcolor{gray!20} $\mathbf{99.42\Rightarrow97.77}$ & $0.18$ \textperthousand  & $1.65\%\downarrow$ & $82.53\%\downarrow$\\
    \hline
    \end{tabular}
    }
    \label{tab: target ntl performance}
\end{table*}

\noindent\textbf{Baselines:} As the first of its kind work addressing malicious probing of pre-trained encoders, there is no prior work for direct comparison with our \MyMethod. 
The closest related work is the SOTA non-transferable learning, including \textit{NTL}~\cite{wang2022non} and \textit{CUTI}~\cite{wang2023model}. For such baseline work, we adopt the pre-trained model, freeze the encoder part, and train a new downstream head on the prohibited target domain to evaluate the baseline model's resistance against malicious probing.\\ 
\noindent\textbf{Datasets:}
Table~\ref{tab:dataset_summary} lists the twelve datasets for our evaluation. 
In addition to the five digits datasets used in the previous work~\cite{wang2022non,wang2023model}, we also assess datasets with larger label spaces, i.e., EMNIST (47 classes) and CIFAR-100 (100 classes).  
By utilizing text-to-image generators, zero-shot \MyMethod is evaluated with more complex datasets.
We test ImageNette~\cite{imagenette} as the source dataset and the military vehicle dataset~\cite{bose_military_vehicles} as the prohibited target\footnote{\url{https://www.kaggle.com/datasets/amanrajbose/millitary-vechiles}} following a practical scenario.
Further, we evaluate the influence of \MyMethod on three admissible domains--ordinary vehicles\footnote{\url{https://www.kaggle.com/datasets/marquis03/vehicle-classification}}, weapons\footnote{\url{https://huggingface.co/datasets/Kaludi/data-csgo-weapon-classification}}, and animals\footnote{\url{https://www.kaggle.com/datasets/alessiocorrado99/animals10/code}}.
\\
\noindent\textbf{Models:}
Our method is assessed using three prevalent DNN architectures: VGG-11~\cite{simonyan2014very}, ResNet-18~\cite{he2016deep}, and Vision Transformer (ViT)~\cite{caron2021emerging}.
We leverage the supervised pre-trained models for VGG-11 and ResNet-18\footnote{\url{https://pytorch.org/vision/stable/models.html}} and fine-tune them on various authorized domains (source).
The early convolutional layers (residual blocks) of VGG-11 and ResNet-18 are considered encoders, while the output dense layer(s) are used as downstream heads for probing.
The ViT utilizes a vision encoder structure that is trained with self-supervised learning.\\
\noindent\textbf{Hyperparameters:} Hyperparameters for supervised and unsupervised \MyMethod can be found in Appendix~\ref{appendix: hyperparameters}.\\
\noindent\textbf{Metric:}
The metric to quantify the encoder's resistance to malicious probing is the \textit{relative} accuracy drop in both the target and the source domains, defined as $\frac{acc_{o}-acc_{m}}{acc_{o}}$, where
$acc_{o}$ is the probing accuracy of the original encoder, and $acc_{m}$ is the one when modified with protection methods.
A higher accuracy drop indicates a strong restriction on given domains.
We expect the accuracy drop in the source domain to be low for preserving the model integrity, while the accuracy drop in the target domain to be high for robustness to malicious probing. 
In Section~\ref{exp: comparison}, we further introduce the Performance Protection Index (PPI) to evaluate the restriction on prohibited domains and the influence on authorized or admissible domains simultaneously for comparison.\\
\noindent\textbf{Platform:} Our implementation uses PyTorch 1.5.0 on Ubuntu 18.04.6 with NVIDIA TITAN RTX.\\
\update{\noindent\textbf{Training cost:} With this experimental setup, training \MyMethod requires between 0.1 and 6 GPU hours, depending on the encoder architecture and dataset size. Detailed training costs for various configurations are provided in Appendix~\ref{sec: complexity}.}

\subsection{Evaluating the Supervised \MyMethod} \label{sec: exp: target ntl}

\begin{figure}[t]
    \centering
    \includegraphics[width=0.88\linewidth]{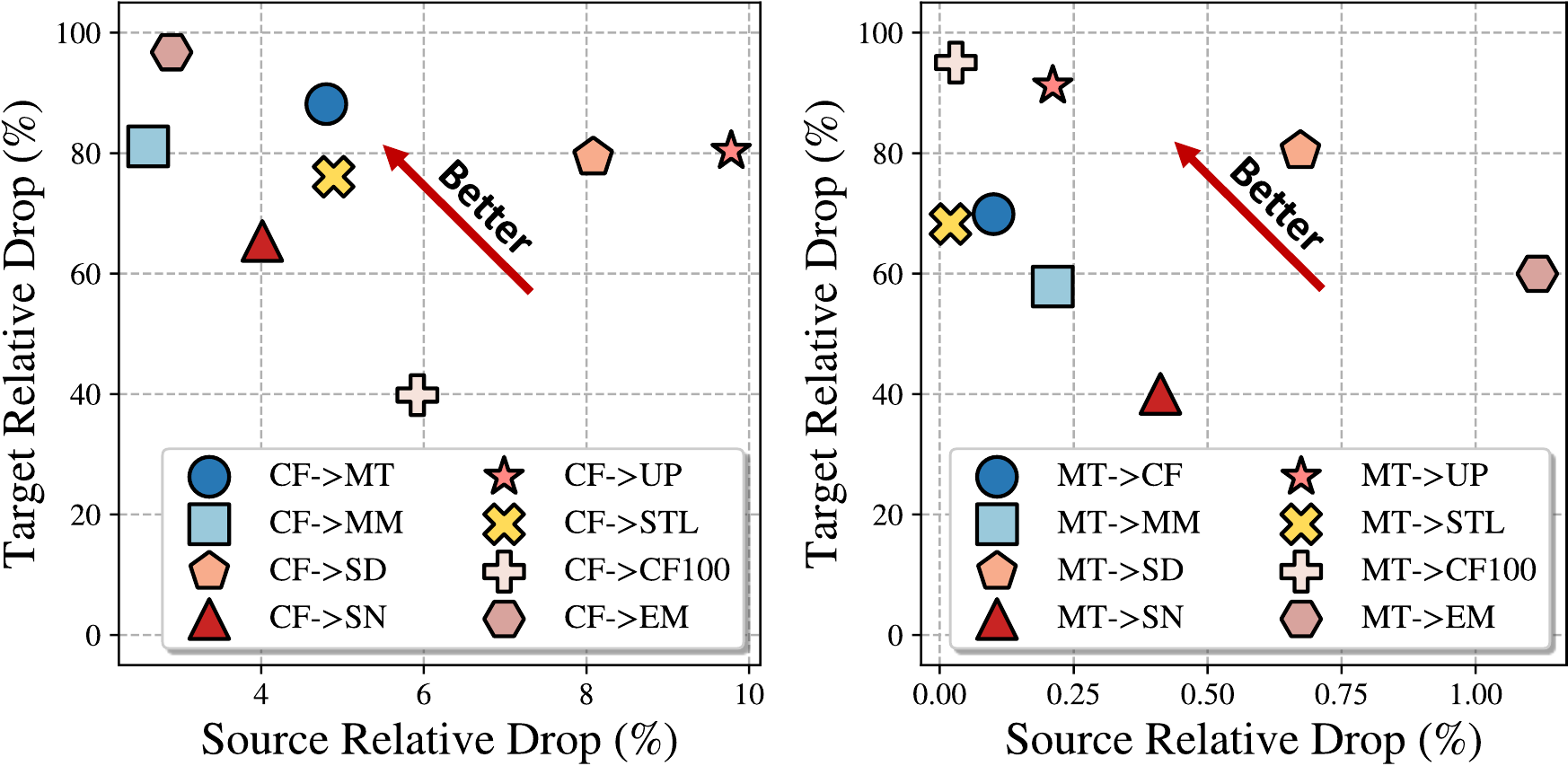}
    \caption{\textbf{Accuracy drop across distinct source and target domains---}It assesses the transferability of VGG-11 encoder with CF (\textbf{Left}) and MT (\textbf{Right}) as the source. Each data point illustrates the simultaneous impact on accuracies on the source and target domain with supervised \MyMethod.}
    \label{fig: accuracy drop scatter}
\end{figure}

\begin{table*}[t]
\setlength{\fboxsep}{1pt} 
    \centering
    \caption{\textbf{\colorbox{darkblue!20}{Unsupervised \MyMethod} performance on encoder (VGG-11) transferability}}
    \resizebox{0.95\linewidth}{!}{
    \begin{tabular}{c|c|c|c|c|c|c|c|c}
    \hline
         \makecell{Source \textbackslash{} Target}& MT & UP & SN & MM & SD & $\Delta W$ & $Drop_\mathcal{S}$ & $Drop_\mathcal{T}$\\
     \hline
    \textbf{MT} &\cellcolor{gray!20}  $\mathbf{99.53\Rightarrow 99.22}$ & $96.35\Rightarrow16.84$ & $43.74\Rightarrow19.61$ & $68.24\Rightarrow12.26$ & $69.65\Rightarrow17.90$ & 0.12\textperthousand  & $0.31\%\downarrow$ & $73.51\%\downarrow$ \\
    \hline
    \textbf{UP} & $97.70\Rightarrow45.02$ & \cellcolor{gray!20} $\mathbf{97.91\Rightarrow 96.44}$ & $58.23\Rightarrow9.59$ & $65.24\Rightarrow13.88$ & $87.10\Rightarrow12.66$ &  0.22\textperthousand  & $1.50\%\downarrow$ & $75.41\%\downarrow$ \\
    \hline
    \textbf{SN}  & $95.30\Rightarrow20.74$ & $92.68\Rightarrow17.09$ & \cellcolor{gray!20} $\mathbf{94.04\Rightarrow94.33}$ & $71.51\Rightarrow50.85$ & $96.96\Rightarrow 94.01$ & 0.21\textperthousand  & $0.31\%\uparrow$ & $47.93\%\downarrow$ \\
    \hline
    \textbf{MM}  & $98.85\Rightarrow40.26$ & $94.67\Rightarrow30.74$ & $53.80\Rightarrow33.97$ & \cellcolor{gray!20}$\mathbf{94.24\Rightarrow93.30}$ & $85.89\Rightarrow55.79$ & 0.15\textperthousand  & 0.99\%$\downarrow$ & $49.68\%\downarrow$ \\
    \hline
    \textbf{SD}& $97.13 \Rightarrow76.68$ & $93.57\Rightarrow86.75$ & $90.40\Rightarrow75.31$ & $71.53\Rightarrow 27.23$ &\cellcolor{gray!20}  $\mathbf{99.83\Rightarrow99.44}$ & 0.20\textperthousand  & $0.39\%\downarrow$ & $26.74\%\downarrow$\\
    \hline
    \end{tabular}
    }
    \label{tab: unsupervised encoderlock performance}
\end{table*}

Table~\ref{tab: target ntl performance} demonstrates the performance of supervised \MyMethod across the datasets of digits for VGG-11 and ResNet-18. 
We compare the accuracy after probing the encoder with corresponding default downstream classification heads, before and after applying supervised \MyMethod, report the relative accuracy drop on the source domain (columns $Drop_\mathcal{S}$) and the average accuracy drop on the target domains ($Drop_\mathcal{T}$), and also present the average percentage of weight change (column $\Delta W$).
The experimental results of VGG-11 reveal that \MyMethod exhibits a steep reduction (up to $78.70\%$) in performance on the target domains while ensuring minimal degradation on the source domain (highlighted in bold, up to $3.53\%$).
Moreover, the accuracy degradation on ResNet-18 shows even a better restriction on the prohibited domain, from $71.73\%$ to $86.02\%$.
In contrast, the degradation of ResNet-18 encoder with \MyMethod on the authorized domain is minimized from $0.28\%$ to $1.65\%$.
Moreover, with less than $0.08\%$ of the weights changed on average, the supervised \MyMethod preserves a higher generalizability of the pre-trained encoder to the admissible domains and avoids the catastrophic forgetting of the encoder's pre-learned knowledge. It will be further discussed in Section~\ref{exp: comparison} as a comparison between different \MyMethod and baseline methods.

\yf{all these observations only apply to VGG-11 results in Table II. What about the ResNet-18 results}
\rd{Add the comparison}

In addition to the accuracy drop, we find that the complexity of the datasets (domains) affects the \MyMethod's performance.  
For instance, on VGG-11, we observe that restricting transferring from a complex domain (e.g., SN, MM, and SD, comprising RGB-colored digit images) to a simple domain (e.g., MT and UP, including grayscale images) is more challenging. 
Specifically, the supervised \MyMethod requires more weights to be changed, and yields a smaller target accuracy drop and a larger source accuracy drop, when the source domain is SN which is a more realistic, 3-channeled digits dataset (the Street View House Number)
A similar phenomenon is also observed in ReseNet-18 supervised \MyMethod, with the highest average weight modification at $0.041\%$.
Moreover, the similarity between the source and target domains also affects the supervised \MyMethod's performance, e.g., the non-transferability of SD$\rightarrow$SN is the worst as they are similar.
\yf{all these observations only appy to VGG-11 results in Table II. What about the ResNet-18 results}
\rd{added}

Unlike prior research~\cite{wang2022non, wang2023model} that only examines the applicability authorization between domains that share the same label space (e.g., 0$\sim$9 digits), during malicious probing, one could aspire to transfer the queried features from the encoder to domains with distinctly different label spaces. Thus, we also evaluate the performance of supervised \MyMethod on VGG-11 under various circumstances: a) transition between distinct task types (CF to MT); b) variation in the size of the label space for similar tasks (CF to CF100); c) changes in both the task type and the label space size (CF to EM).
The results are presented in Fig.~\ref{fig: accuracy drop scatter}. Overall, supervised \MyMethod achieves good performance across all these transfer tasks, manifesting a much higher drop in the target domain than the source. Similar to experiments on digit datasets, the more distinct the target dataset is from the source dataset, the better performance. 
For instance, supervised \MyMethod performs well when transferring across different tasks, i.e., transitioning from digits to images (CF) and vice versa; while it results in a lower performance when transferring from CF to CF100 as these two share a large number of similar features.
Notably, when MT is the source domain and CF100 is the target domain, \MyMethod can significantly reduce the accuracy on CF100 to $1.19\%$, closely resembling random guessing across the $100$ classes, i.e., the encoder fails to extract features. 
While for transferring between MT and EM (both digits), the accuracy only drops to $32.04\%$. 
We posit that a pre-trained encoder tends to capture more intricate features when applied on similar domains, while there will be more distinct features when the source and target domains diverge significantly. We characterize the similarity between datasets using MMD in Appendix~\ref{appendix: dataset similarity}.

\subsection{Evaluating the Unsupervised \MyMethod} \label{exp: unsupervised-encoderlock}

\begin{figure}[t]
    \centering
    \includegraphics[width=0.92\linewidth]{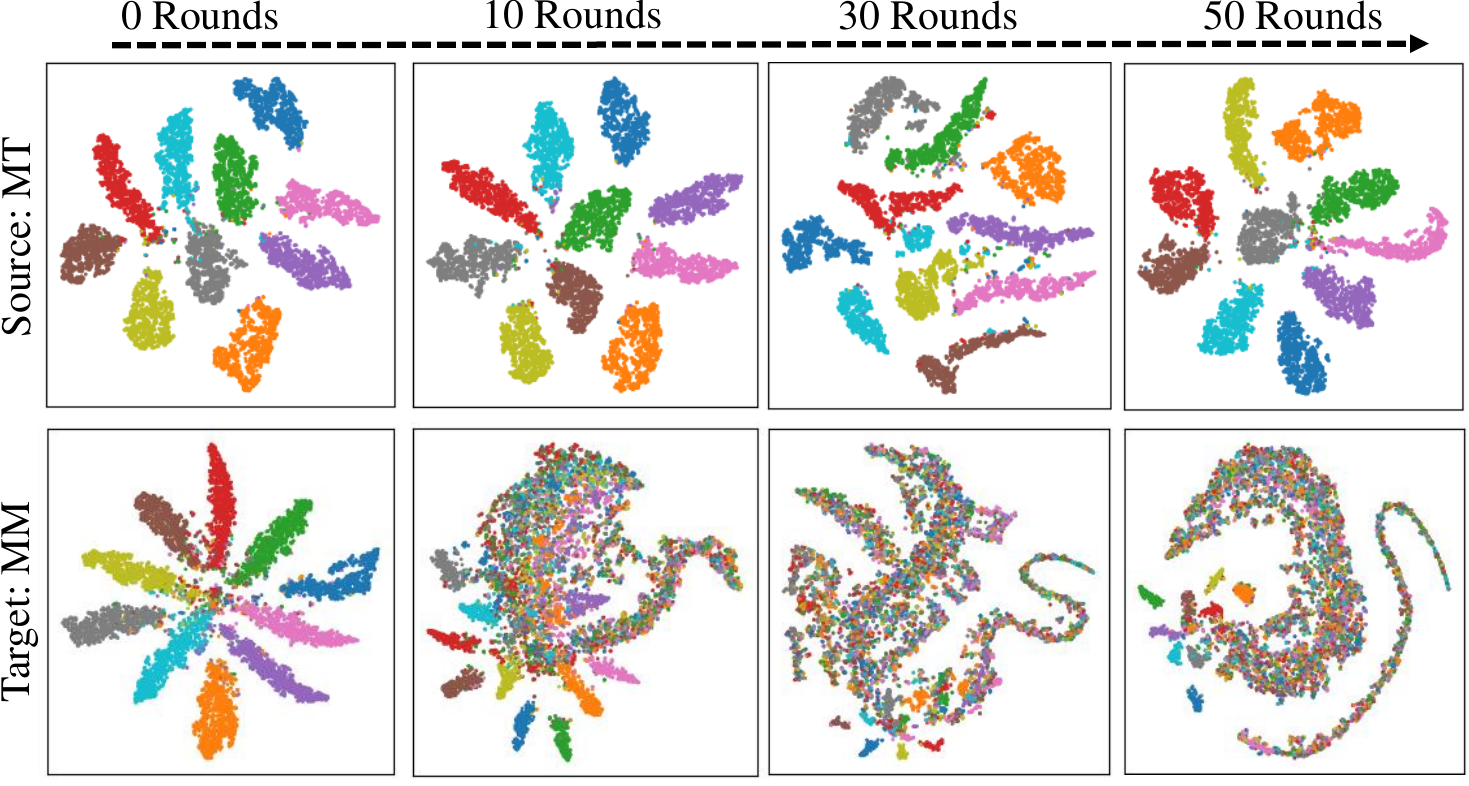}
    \caption{\textbf{Unsupervised \MyMethod}-Latent Space Change via Rounds}
    \label{fig: unsupervised-latent space}
\end{figure}
The unsupervised \MyMethod addresses the challenge when the model owner has no access to the true labels of samples from the target domain.
To demonstrate the effectiveness of the proposed contrastive loss in Eq.~(\ref{eq: unsupervised loss}), we conduct the same evaluation on digits datasets with results on VGG-11 shown in Table \ref{tab: unsupervised encoderlock performance} and ResNet-18 in Appendix~\ref{appendix: unsupervised resnet18} Table~\ref{tab: unsupervised encoderlock performance-resnet}.
The unsupervised \MyMethod also successfully restricts the encoder's transferability to the target domain and maintains the encoder's integrity on the source domain, with a small number of weight changes.

Compared with supervised \MyMethod, the unsupervised \MyMethod demonstrates better performance in preserving accuracy within the source domain, but worse performance in reducing the target accuracy. 
This is due to the proposed $R^{\text{cont}}_\mathcal{T}$, utilizing contrastive loss, necessitates a more discriminative latent space for the source domain by reducing $L^{\text{cont}}_\mathcal{S}$.
To further illustrate, in Fig.~\ref{fig: unsupervised-latent space} we visualize the change of the latent space from both the source (MT) and target (MM) domains using t-SNE~\cite{van2008visualizing}. With more rounds of unsupervised \MyMethod (indicating more training epochs and more weight updates), the source domain remains class-discriminative, while the target domain becomes less discriminative. 
However, some small clusters of classes can still be observed, which leads to a higher accuracy when we fine-tune a downstream classifier on such a latent space on the target domain.

\subsection{Evaluating the Zero-shot \MyMethod} \label{exp: zero-shot encoderlock}

\begin{figure*}[t]
    \centering
    \includegraphics[width=0.95\linewidth]{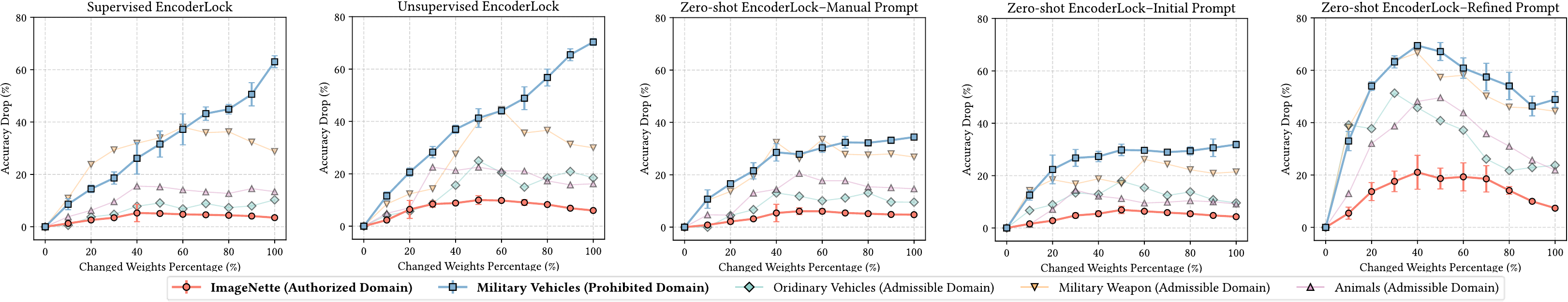}
    \caption{\textbf{Comparison among different \MyMethod}--on authorized, prohibited, and admissible domains. We randomly select a percentage of domain-aware weights as the X-axis, indicating the optimization process of domain-aware weights. The authorized and prohibited domains are run $5$ times with error bars plotted.}
    \label{fig: different-encoderlock-performance}
\end{figure*}

For zero-shot \MyMethod, we utilize the GPT-4 API as the AI agent to interpret the semantic meaning of the prohibited theme\footnote{\url{https://platform.openai.com/docs/models/gpt-4-turbo-and-gpt-4}} with a pre-trained stable diffusion model for text-to-image generation\footnote{\url{https://huggingface.co/CompVis/stable-diffusion-v1-4}} to create synthetic datasets.
To show the effectiveness of the AI agent and prompts refining (Fig.~\ref{fig: zero-shot-framework}), we test the zero-shot \MyMethod under three scenarios: 
1) The prompts are created \textit{manually} (without AI agents); 
2) An AI agent generates prompts \textit{randomly};
3) The AI agent \textit{generates} prompts and \textit{refines} them.
We utilize 10 prompts to the generative model for 1,000 fake images as the prohibited dataset. 
Prompts and synthetic images can be found in Appendix~\ref{appendix: synthetic images}.

We evaluate one-shot \MyMethod's performance on the source domain (ImageNette~\cite{imagenette}) and the target domain (a real military vehicle dataset~\cite{bose_military_vehicles}) with a pre-trained encoder using ResNet-18.
Furthermore, we select three admissible domains to evaluate \MyMethod's impact on other domains, neither source nor target.
Considering the semantics of `military vehicles', we utilize two admissible datasets with closer semantic meaning and one unrelated.
1). \textbf{Ordinary vehicles}: A 10-class normal vehicle classification dataset related to the prohibited domain in `vehicle', i.e., sedan, SUV, and bus.
2). \textbf{Weapons}: An 11-class game-based weapon classification dataset, related to the prohibited domain in `military', i.e., AK-47, Famas, and UPS.
3). \textbf{Animals}: A 10-class animal classification dataset with no obvious semantic meaning with the prohibited domain, i.e., chicken, cat, and butterfly.

Fig.~\ref{fig: different-encoderlock-performance} shows the performance of different \MyMethod variants on authorized, prohibited, and three admissible domains.
We present the testing accuracy degradation on each domain versus the percentage of critical weights modified.
The original pre-trained encoder has an average accuracy $96.59\%$ on the authorized domain and $60.55\%$ on the prohibited domain.
Supervised \MyMethod shows the strongest performance restriction on the prohibited domain ($11.48\%$) and meanwhile preserving a high accuracy on the authorized domain ($94.17\%$). 
The primary reason behind the efficacy of the supervised \MyMethod is its well-defined prohibited domain. This clarity allows for targeted modifications of critical weights. As a result, \MyMethod can enhance specific aspects of the encoder's performance while maintaining its overall generalization capability.
Unsupervised \MyMethod demonstrates the second highest performance with a label-free prohibited domain by restricting the target performance to $18.34\%$ and keeping the accuracy on the source at $93.43\%$.
The introduction of contrastive loss penalizes more general features than the supervised loss, causing a larger degradation on the authorized and admissible domains.
Zero-shot \MyMethod works on the most challenging accessibility to the prohibited domain.
Without any data from it, using prompts from military fans (manual prompts in Appendix~\ref{appendix: synthetic images}) shows similar to the one-time initial prompt from the AI agent. 
The prompts and generated synthetic images are not general enough to describe the entire prohibited domain, thereby leading to the smallest restriction on the military vehicle dataset. (Manual: $41.81\%$, AI initial: $38.61\%$)
When we refine the AI prompts for more general features of the prohibited domain with the presented refinement algorithm, there is a noticeable increase in the \MyMethod restriction ability ($23.69\%$).
However, such prompts cause the largest accuracy degradation on the authorized domain ($92.86\%$), as its wide restriction on the encoder's features.
\update{In addition to the relevance of prompts, the synthetic dataset's quality also affects the strength of \MyMethod.
Specifically, a high-quality synthetic dataset, generated after large number of inference iterations and bearing small noise, enhances the defense capability of \MyMethod. 
More detailed results are shown in Appendix~\ref{appendix: zero-shot quality}.}

Other than the performance of authorized and prohibited domains, our evaluation on the admissible domains indicates the impact of \MyMethod on the generalizability to unknown data.
In particular, all five \MyMethod's restrictions on the prohibited `military vehicles' show higher penalties on the `weapon' dataset.
On the other hand, the impact on the encoders' performance on ordinary vehicles is small, similar to what from the animal dataset.
We will further present a deep analysis of this phenomenon in Section~\ref{sec: discussion: visualization}, where we show the encoder's attention shifting from the attack module of the military vehicles, i.e., the barrel.
As a result, it has a lower effect on distinguishing between various types of vehicles but has a significant impact on different types of firearms.

In Fig.~\ref{fig: different-encoderlock-performance}, we also present the probing performance of \MyMethod when the portion of critical weights to change varies.
In particular, the pre-trained encoder's performance keeps dropping on the prohibited domain but its performance degradation on the authorized and admissible domain shows a `degradation peak' at the point of changing half of critical weights.
This phenomenon is likely due to the integrity of \MyMethod's critical weights updating process affected by the random sampling.
A complete updating process with either supervised loss~\eqref{eq: loss} or unsupervised loss~\eqref{eq: unsupervised loss} ensures the encoder cannot perform well on the target domain but still effective on the source domain, as shown the good performance when changing $100\%$ critical weights.
However, only updating half of them will break the encoder's integrity on the source domain and the admissible domain as it breaks the network connectivity between critical weights, leading to unstable \MyMethod.
Such phenomenon also shows in the standard deviation on the curves in Fig.~\ref{fig: different-encoderlock-performance}, where when only changing half of critical weights, the \MyMethod performance has the largest variance.

\yf{not clear about the reason.particularly what is the "pathway". rephrase it} 
\rd{revised}
\subsection{Comparison with Prior Methods} \label{exp: comparison}
\begin{figure*}[t]
    \centering
    \includegraphics[width=0.96\linewidth]{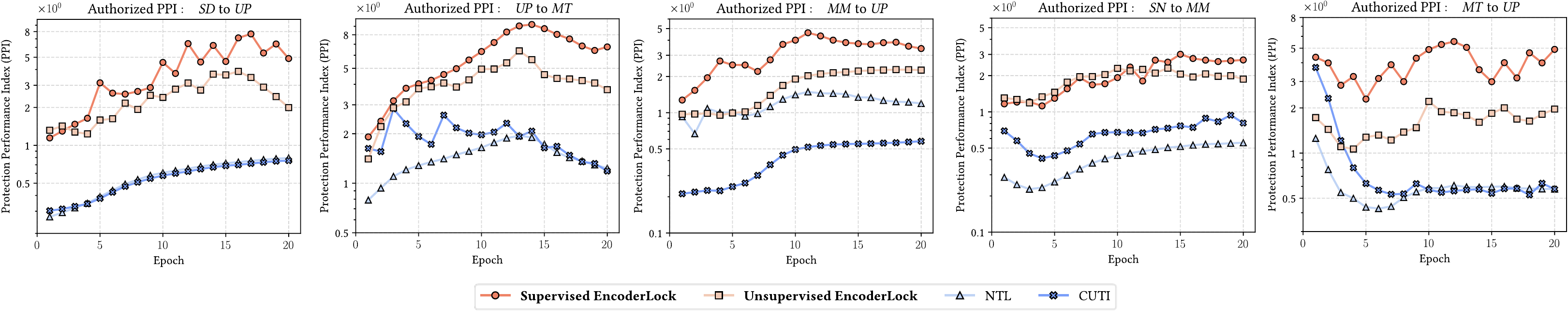}
    \caption{\textbf{Comparison on Authorized PPI}--between \MyMethod with NTL~\cite{wang2022non} and CUTI~\cite{wang2023model} on different pairs of authorized domains and prohibited domains. \textbf{The higher the better.} Probing for multiple epochs can't increase the performance on prohibited domains but keeps its performance on authorized domains.}
    \label{fig: authorized applicability score}
\end{figure*}

\begin{figure*}[t]
    \centering
    \includegraphics[width=0.96\linewidth]{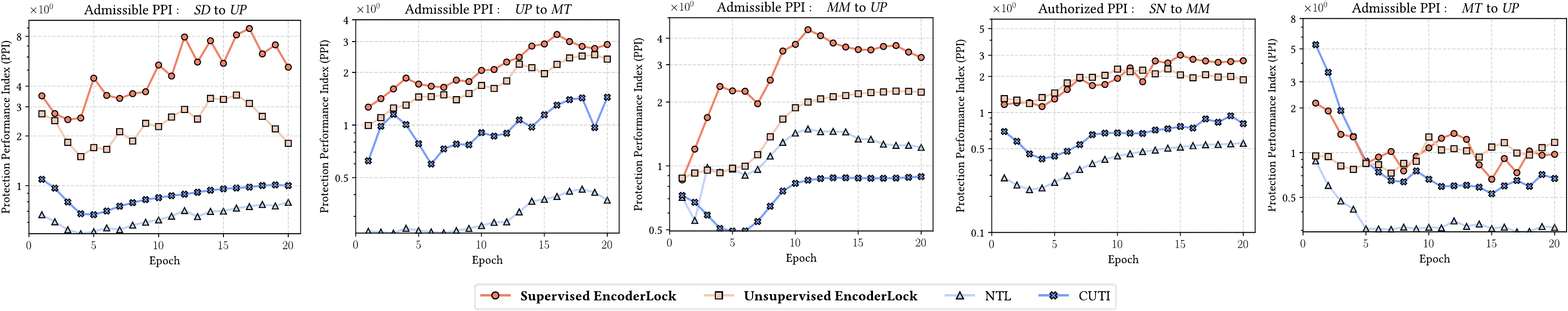}
    \caption{\textbf{Comparison on Admissible PPI}--between \MyMethod with NTL~\cite{wang2022non} and CUTI~\cite{wang2023model} on different pairs of admissible domains and prohibited domains. \textbf{The higher the better.} \MyMethod shows the minimal impact on admissible domains while restricting the encoder's performance on prohibited domains.}
    \label{fig: admissible applicability score}
\end{figure*}

We compare \MyMethod with SOTA baselines: NTL~\cite{wang2022non} and CUTI~\cite{wang2023model} in protection against malicious probing. 
Specifically, we probe the encoder on the target dataset with fine-tuned downstream heads. 
Moreover, we propose a new stable metric--\textit{Protection Performance Index (PPI)}--to measure the performance in restricting on the target domain ($\mathcal{D}_\mathcal{T}$) relative to what in retention on the base domain ($\mathcal{D}_\mathcal{B}$), defined as:
{\small
\begin{equation}
    \text{PPI}(\mathcal{D}_\mathcal{T}, \mathcal{D}_\mathcal{B}) = \frac{acc_{o}^{\mathcal{T}}/acc_{m}^{\mathcal{T}}}{acc_{o}^{\mathcal{B}}/acc_{m}^{\mathcal{B}}}
\end{equation}}PPI measures the ratio of change in performance before and after applying protection for a prohibited target domain with an authorized or admissible base domain. 
A higher PPI indicates better non-transferability and resistance to malicious probing.

In Fig.~\ref{fig: authorized applicability score}, we compare the PPI of supervised and unsupervised \MyMethod with the baselines in five pairs of authorized and prohibited domains when probing the downstream heads by epochs.
Our proposed supervised \MyMethod shows the best non-transferability performance, while the unsupervised \MyMethod also demonstrates good performance overall. 
In addition, we also present the PPI between admissible domains and prohibited domains in Fig.~\ref{fig: admissible applicability score}.
The PPI of supervised and unsupervised \MyMethod is still better than the baselines.
\yf{what do you mean slightly better? what is the range for much better and slightly better respectively} \rd{revised}
However, the advantage is not obvious due to the prohibited and admissible domains are highly similar as they are all digital datasets.
Therefore, critical weights on prohibited domains often overlap with those in admissible ones, causing a higher accuracy drop as the weights are not optimized during the updating process of the \MyMethod.
A good example can be found in Fig.~\ref{fig: different-encoderlock-performance}, where we analyze the admissible domain with distinct semantic meaning (i.e., animals versus military vehicle), \MyMethod can preserve the encoder generalizability to semantic unrelated domains.
Additional numerical results can be found in Appendix~\ref{appendix: comparison} Table~\ref{tab: comparison with baseline methods-appendix-1} to \ref{tab: comparison with baseline methods-appendix-8}.

\begin{figure}[t]
    \centering
    \includegraphics[width=0.88\linewidth]{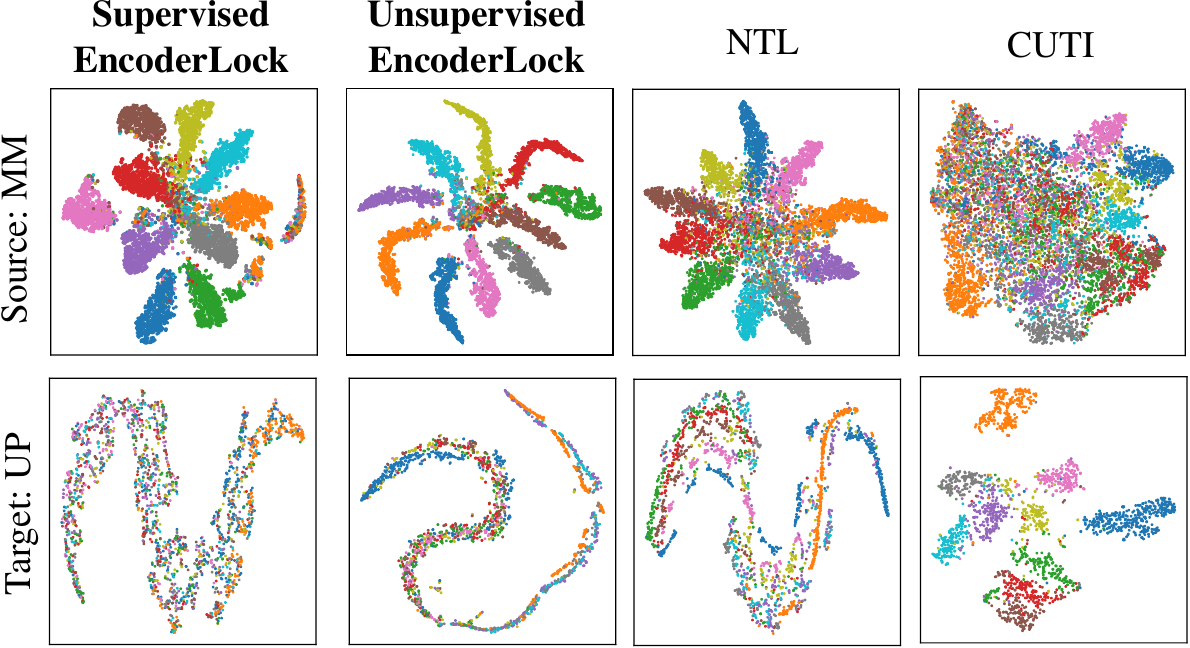}
    \caption{\textbf{Latent Feature Visualization}--Feature Space of Different Methods}
    \label{fig: baseline-latent-space}
\end{figure}

When we fine-tune the downstream classifier on the encoder trained using NTL and CUTI, the encoder performance on the source domain may decrease considerably, or its performance on the target domain may still be high. 
This indicates that previous methods are not suitable for the scenario of malicious probing.
To explain this phenomenon, we visualize the source (MM) and target (UP) domains on the latent space with different trained encoders in Fig.~\ref{fig: baseline-latent-space}, where the colors represent the true labels of the testing dataset. The self-challenging scheme employed in supervised \MyMethod effectively preserves a higher discrepancy within the source domain, while very little class-related information is discernible within the target domain. The unsupervised \MyMethod, despite lacking class information, still successfully reduces class distinguishability on the target domain.
In contrast, NTL~\cite{wang2022non} penalizes the Maximum Mean Discrepancy between the source and target latent spaces, showcasing the distribution difference between these domains, but samples from the same class still cluster together in the target domain. Therefore, its transferability can be resumed via fine-tuning. CUTI~\cite{wang2023model} only considers target performance during training and shows the worst non-transferability--its target domain is clearly clustered by classes.

\subsection{Interpretation of \MyMethod} \label{sec: discussion: visualization}

\begin{figure*}[t]
    \centering
    \includegraphics[width=0.86\textwidth]{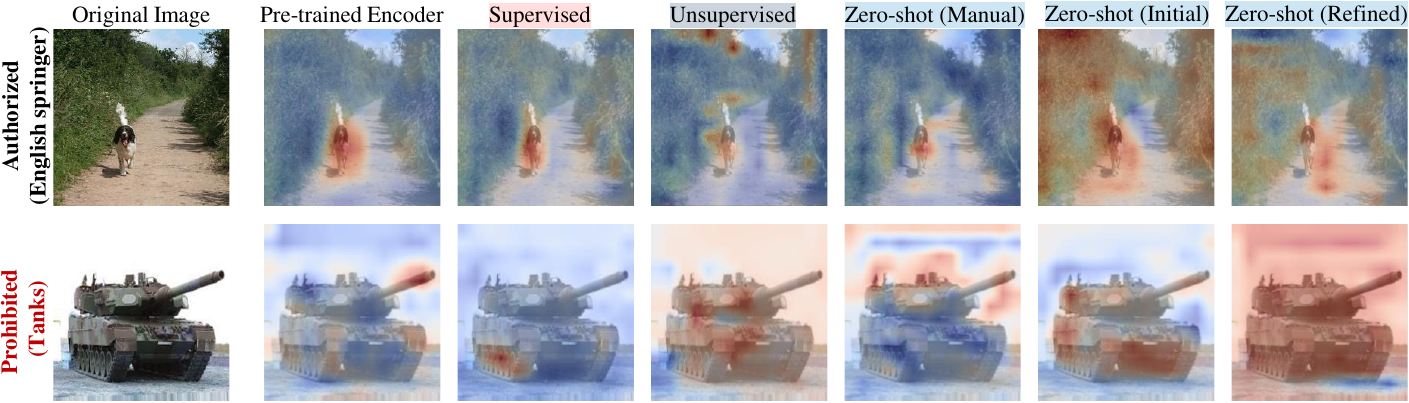}
    \caption{\textbf{Interpretation of Different \MyMethod using GradCAM~\cite{selvaraju2017grad}}--the {\color{red} red} parts highlight the focus of encoder to make decisions.}
    \label{fig: visual interpretation}
\end{figure*}

To further understand the changes in the encoders generated by different variants of \MyMethod, we use the encoders trained from Section~\ref{exp: zero-shot encoderlock} and visualize their decision-making process with Gradient-weighted Class Activation Mapping (GradCAM)~\cite{selvaraju2017grad}, as illustrated in Fig.~\ref{fig: visual interpretation}. Specifically, the GradCAM attribution is computed for the last convolutional layer in the encoder (ResNet-18) and is upsampled to act as a mask added to the original input (the cool-warm heatmap in Fig.~\ref{fig: visual interpretation}). The highlighted red part indicates the feature that the encoder focuses on to make its prediction.
We observe that the original pre-trained encoder focuses on the English Springer correctly, and it performs well on the military dataset (tanks) as its focus is moved to the main gun barrel of the tank. However, the supervised \MyMethod, which has less effect on the source domain data, switches the encoder's focus to the tank track, leading the fine-tuned downstream classifier to make a wrong prediction as `Armored combat support vehicles'.
The unsupervised \MyMethod, aiming to blur the entire feature space's class-discrepancy, generates a more vague interpretation of the decision process—the focus is mainly on the vehicle but not on specific features of tank.
In Appendix~\ref{append: interpretation admissible}, we also visualize the GradCAM results on three different admissible domains. 
The findings indicate that the attack module of military weapons also experiences a loss of focus in the model, resulting in a significant impact of \MyMethod on the weapon dataset.

We also visualize the GradCAM of zero-shot \MyMethod with different types of prompts. The manual prompts show less effectiveness in the model's non-transferability, still focusing on some useful features such as the tank roof and gun barrel. The AI-agent-generated prompts (with or without refinement) have a strong effect on the source domain, especially the green part (bash in the source figure), considering `green' is likely related to the camouflage. With the proposed prompt refinement process, the synthetic dataset obfuscates most of the useful features of the military theme, leading to a very vague focus on the target domain.
The interpretation of \MyMethod further reinforces its effectiveness: the supervised \MyMethod has the highest performance as it has the ground truth labels, and therefore able to move away the encoder's focus on specific features; while the unsupervised and zero-shot \MyMethod directly cause the latent feature space to be less informative, leading to all input features having similar importance.

\subsection{Real-world Case Study} \label{sec: case-study}

Previous evaluations mainly focus on small encoders extracted from supervised-trained DNN models with basic architectures (e.g., VGG-11 and ResNet-18).
To demonstrate that \MyMethod is practical in protecting real-world encoders, we apply it to a public encoder based on Vision Transformer (ViT) released by Facebook\footnote{\url{https://huggingface.co/facebook/dino-vits16}}\cite{caron2021emerging}. 
This encoder is a transformer pre-trained on a large collection of images (ImageNet-1k~\cite{deng2009imagenet} with a resolution of $224$$\times$$224$) in a self-supervised fashion.
The input images are presented to the model as a sequence of fixed-size patches (the patch size is $16$$\times$$16$).
ViT learns an inner representation of images that can extract features useful for downstream tasks with probing heads (downstream models).
As reported, the training process requires approximately 2.6 days with a computational power of 16 GPUs. 
The pre-trained encoder represents a valid IP, and our proposed \MyMethod for applicability authorization aims to protect the IP.

\begin{figure}[t]
    \centering
    \includegraphics[width=0.95\linewidth]{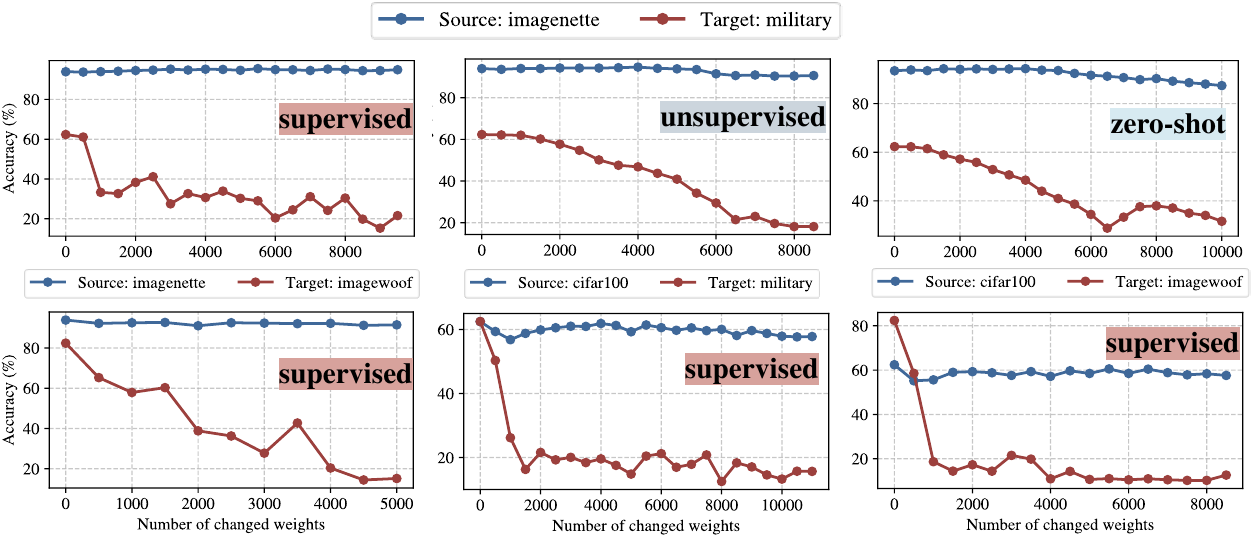}
    \update{
    \caption{\textbf{Evaluation on a Pre-trained ViT}: \textbf{Top row} - source (target) accuracy versus number of changed weights for different levels of \MyMethod;  \textbf{Bottom row}: performance of supervised \MyMethod on different datasets.}    \label{fig: vit-results}
    }
\end{figure}

We first evaluate the performance of the supervised \MyMethod, considering ImageNette~\cite{imagenette} and CIFAR-100~\cite{radford2021learning} as the source domains and the military dataset and Imagewoof dataset as the target domains. The ImageWoof dataset is a selected subset of ImageNet with different types of dogs\footnote{\url{https://github.com/fastai/imagenette}}. Note that to fit the input resolution of ViT, the CIFAR-100 inputs are resized to $224$$\times$$224$. We also evaluate the ViT with \MyMethod on some simple datasets, as shown in Appendix~\ref{appendix: vit on simple datasets}.
In Fig.~\ref{fig: vit-results}, we visualize the accuracy after fine-tuning the downstream classifier on the source and target domains, respectively, versus the number of weight changes. We observe that for the pre-trained ViT, the supervised \MyMethod still works effectively in restricting the target domain performance and preserving the source domain performance. In addition, the largest accuracy drop often happens in the early epochs (early sets of domain-specific weights). This demonstrates the effectiveness of the proposed domain-aware weight selection algorithm when the most important (target-domain sensitive) weights are selected.

\update{Similarly, we apply unsupervised and zero-shot \MyMethod on the same ViT encoder between the ImageNette (i.e., authorized) and military (i.e., prohibited) datasets, 
and the results are shown in Fig.~\ref{fig: vit-results}.
All three variations of \MyMethod effectively limit the encoder's performance on the prohibited domain. Notably, supervised ($21.56\%$) and unsupervised ($18.15\%$) \MyMethod provide stronger protection than the zero-shot version ($29.26\%$), due to their higher accessibility during training.
And supervised \MyMethod outperforms in maintaining the accuracy in the authorized domain, as the more accurate penalization on the labeled prohibited domain.
In conclusion, all three variants of \MyMethod demonstrate effectiveness for safeguarding real-world pre-trained encoders.}

\yf{why the target domain accurcy starts bad for the first case of "imagenette to military" and the performance drops in the target domain only with a lot of weight changes?}
\rd{I suspect the features from imagenette is more general so the original probing performance is not high, and we need more modification to restrict its useful information on the military dataset. Cifar100 is more simple, so the encoder has a worse generalizability to military, therefore, we only need few weights to break its performance. }

\section{Discussions} \label{sec: discussion}
\begin{table*}[t]
\caption{\textbf{Performance of the supervised \MyMethod on various classifier configurations---}source(MT) to target(UP).}
\centering
\resizebox{0.95\linewidth}{!}{
\begin{tabular}{c|c|ccccc|ccccc|ccccc}
\hline
\# Layers & 1 & \multicolumn{5}{c|}{2}          & \multicolumn{5}{c|}{3}          & \multicolumn{5}{c}{4}          \\  \hline
Hidden dim.    & /  & 256 & 512 & 1024 & 2048 & 4096 & 256 & 512 & 1024 & 2048 & 4096 & 256 & 512 & 1024 & 2048 & 4096\\

\hline
size (M) & 0.25 &   6.42 & 12.85 & 25.7 & 51.4 & 102.8 & 6.49 & 13.11 & 26.75 & 55.59 & 119.59 & 6.56 & 13.38 & 27.8 & 59.8 & 136.37 \\
  
$Acc_{\mathcal{S}}^{el}$ (\%)    &  99.12 &     99.09 & 99.21 & 99.20 & 99.29 & 99.20 & 99.32 & 99.29 & 99.24 & 99.27 & 99.36 & 99.29 & 99.29 & 99.26 & 99.19 & 99.30 \\

$Acc_{\mathcal{T}}^{el}$ (\%)  & 17.89  &   17.89 & 17.89 & 13.15 & 17.89 & 17.89 & 9.87 & 17.89 & 17.89 & 17.89 & 17.89 & 13.15 & 17.89 & 8.47 & 17.89 & 17.89 \\
 \hline     
\end{tabular}}
\label{tab:black-box}
\end{table*}

\subsection{Ablation Studies of \MyMethod} \label{sec: ablation}
\subsubsection{\MyMethod on Probing with Various Architectures} \label{sec: exp: unknown classifier}
As shown in Challenge 2, the attacker takes full control of the downstream classifier and can optimize it for malicious probing. 
In this section, we further evaluate the protection performance of \MyMethod with various downstream head structures.
Taking supervised \MyMethod on VGG-11 as an example, we delve into diverse depths and widths for the linear probing heads, including varying numbers of layers and hidden dimensions  (no hidden layer when there is only one layer in the classifier).
The detailed results of these experiments are presented in Table~\ref{tab:black-box}, which shows that
one can leverage the features from our pre-trained encoders to achieve impressive performance ($>99\%$) on the source task but only achieve up to $17.89\%$ on the target task, across a comprehensive collection of configurations of the downstream classifiers. 
This demonstrates the effectiveness of the self-challenging training scheme. 
\yf{are these all configurations for classifier? does it have to be fully-connected? any other architecture options?} \rd{For probing, people are often use linear head. I think we main focus on multiple depths, width of these linear heads.}

\yf{what is ``novel black-box defense"? it is basically your threat model - the owner does not know how the user is going to optimize the downstream classifiers.  he has to sort of "simulate"/optimize future downstream classifiers.} \rd{Yes, we do not know the architecture of downstream heads as well, this section is focused on the structure difference.}

\update{
\subsubsection{\MyMethod on Probing with More Prohibited Data} \label{sec: data amount}
We consider a real-world attacker who can be adaptive in attacking the encoders. In our initial malicious probing scenario, we assume an attacker uses a small amount of data ($10\%$) to attempt redirecting the pre-trained encoder. However, once the attacker becomes aware of the defense mechanism, s/he can probe the downstream head with a larger amount of prohibited data. We assess the resistance of \MyMethod against such possible attacks in Appendix~\ref{appendix: various data}.
Thanks to the self-challenging training in supervised \MyMethod and the manipulation of the latent feature space in unsupervised \MyMethod, our approach demonstrates strong resistance to malicious probing, even when the attacker gains access to the entire  prohibited dataset.
}

\begin{figure}[t]
    \centering
    \includegraphics[width=0.98\linewidth]{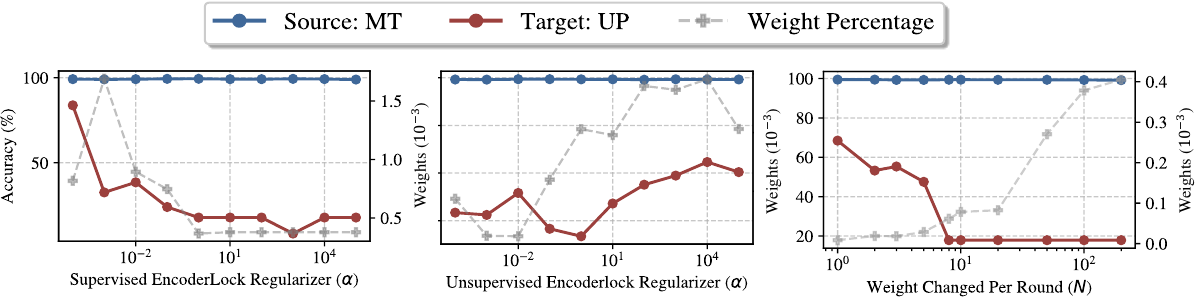}
    \caption{\textbf{Ablation studies--}a) regularizer weights for \colorbox{koblue!20}{supervised \MyMethod}  $\alpha$; b) regularizer weights for \colorbox{kopink!20}{unsupervised \MyMethod};c) changed weights number per round(N);
    The probing accuracy on the source and target domains are reported on the \textbf{left} y-axis and the changed weight percentage is on the \textbf{right} y-axis. Original accuracy on source and target are \colorbox{darkblue!30}{$99.53\%$} and \colorbox{koblue!30}{$94.91\%$}.}
    \label{fig: ablation on data volumn}
\end{figure}

\subsubsection{\MyMethod with Various Hyperparameters}
we employ the transfer scenario from MT to UP as an example.\\
\noindent\textbf{Regularization Term ($\alpha$):}
The regularization term in Eqn.~(\ref{eq: regularization}) plays an important role in balancing the model integrity on the source domain and preventing malicious probing on the target domain. We vary the weight $\alpha$ and Fig.~\ref{fig: ablation on data volumn} (a) (b) show the effect of the regularization term for supervised and unsupervised \MyMethod, respectively.
It is worth noting that the regularization term ensures that all $\alpha$ values sustain the original performance on the source domain, due to the log-scale term.
Comparing different $\alpha$, we choose $\alpha=10^3$ for supervised \MyMethod and $\alpha=1$ for unsupervised \MyMethod.
\\
\noindent\textbf{Number of Critical Weights to Update per Round ($N$):} \lls{inconsistency with the input list of your function 1
}
Fig.~\ref{fig: ablation on data volumn} (c) inspects the repercussions of varying the number of critical weights (\textit{N}) selected in each round, in the range of $1$ to $200$. 
When \textit{N} is too small, the training of \MyMethod is slow and can not converge even with the maximum number of rounds ($R=100$). 
A larger \textit{N} will cause more weight change for \MyMethod, reducing the encoder's generalizability.
Our selection of $N$ can be found in Table~\ref{tab: appendix: hyperparameters}.
\update{
\subsection{Security Analysis of \MyMethod} \label{sec: security analysis}

The security of an encoder protected with \MyMethod can be assessed by setting an accuracy threshold ($acc_{th}$) on the prohibited domain, with the accuracy drop of it on the authorized domain below a certain constraint ($\epsilon$).
For example, $acc_{th}$ can be defined by the accuracy of a \textit{train-from-scratch} model, and the $\epsilon$ is set at $2\%$,  as outlined in Appendix~\ref{appendix: security analysis}. In this case, the encoder is deemed secure because an attacker lacks incentive for malicious probing, with its performance no better than direct training.
}

\subsection{Future Works} \label{sec: future works}
In this work, we have demonstrated the effectiveness of \MyMethod for different levels of domain data accessibility. However, \MyMethod still requires the EaaS provider to clearly specify the prohibited domain, meaning that the provider should know what the encoder is allowed to do and what should be prohibited.
This causes inconvenience in automatic detection and restriction on any unknown `theme' of harmful tasks.
One potential avenue for future work is to incorporate \MyMethod with toxicity content detection utilizing large language models~\cite{phute2023llm, wang2022toxicity} to automatically identify and restrict prohibited domains without relying on explicit specifications from the model owner. 
This approach would further enhance the flexibility and adaptability of \MyMethod.
Furthermore, while we have showcased the effectiveness of \MyMethod in image classification tasks, pre-trained encoders are useful for many other applications, such as image generation and semantic segmentation. 
Extending \MyMethod to these different tasks may unlock additional potential in controlling the encoder's transferability. 
The main challenge will be how to combine different forms of loss terms (i.e., generation loss or segmentation loss) with the loss for a pre-trained encoder.
\section{Conclusions} \label{sec: conclusion}
In this work, we address a new security issue arising during the probing process of pre-trained encoders: restricting the applicability to harmful prohibited domains.
We recognize a realistic challenge about the data accessibility to prohibited domains and propose supervised, unsupervised, and zero-shot \MyMethod for different levels of knowledge of prohibited domains. We propose novel and effective domain-aware weight selection and self-challenging training to maintain the encoder's integrity while protecting it against malicious probing.
Our evaluation has validated the efficacy of \MyMethod in resisting malicious probing across various domains and encoders.

\bibliographystyle{IEEEtranS}
\bibliography{sample.bib}

\appendices

\section{Hyperparameter Configuration of \MyMethod} \label{appendix: hyperparameters}
\begin{table}[h]
    \centering
    \caption{Hyperparameters used in the experiment}
    \begin{tabular}{|c|c|c|c|c|}
    \hline
    Hyperparameters& $N$ & $R$ & $\alpha$ & LR \\
    \hline
    Supervised \MyMethod  & $100$ & $100$ & $1,000$ & $0.01$\\
    \hline
    Unsupervised \MyMethod  & $200$ & $100$ & $10$ & $0.01$\\
    \hline
    \end{tabular}

    \label{tab: appendix: hyperparameters}
\end{table}
In this section, we provide the hyper-parameters configuration in this work.
During training \MyMethod, we use $1,000$ samples from the training dataset.
During fine-tuning of the downstream classifiers, we use $10\%$ of the training data and evaluate with the entire testing data (usually $20\%$ of the data).
The fine-tuning process uses the Adam optimizer~\cite{kingma2014adam} with adaptive learning rate scheduling~\cite{zeiler2012adadelta} and an early stopping criterion~\cite{prechelt2002early} (patience=$10$).
For the default supervised \MyMethod configuration, we use $N=100$, $R=100$, and $\alpha=10^3$. 
For the default unsupervised \MyMethod configuration, we use $N=200$, $R=100$, and $\alpha=10^1$. Hyperparameters are evaluated in Section~\ref{sec: ablation}.

 \section{Dataset Similarity} \label{appendix: dataset similarity}
\begin{table}[H]
    \centering
        
            \caption{    \textbf{Datasets' feature space cosine similarity}}
            \resizebox{0.9\linewidth}{!}{
    \begin{tabular}{cccccccccc}
    
    \hline
         &  MT&  UP&  MM&  SN&  SD&  EM&  CF10&  STL10& CF100\\
    \midrule
         MT&  0.999& \textbf{ 0.707}&  0.577&  0.452&  0.706& \textbf{ 0.942}&  0.404&  0.303& 0.448\\
         UP&  \textbf{0.712}&  0.999&  0.551&  0.652&  0.892&  0.786&  0.395&  0.276& 0.425\\
         MM&  0.579&  0.570&  0.993&  0.570&  0.606&  0.590&  0.505&  0.549& 0.5151\\
         SN&  0.467&  0.650&  0.570&  0.998&  0.777&  0.521&  0.522&  0.455& 0.548\\
         SD&  0.712&  0.891&  0.587&  0.768&  0.999&  0.791&  0.452&  0.345& 0.483\\
         EM&  \textbf{0.938}&  0.780&  0.553&  0.514&  0.794&  0.999&  0.404&  0.296& 0.432\\
         CF10&  0.408&  0.398&  0.499&  0.519&  0.454&  0.405&  0.996&  0.831& 0.972\\
         STL10&  0.302&  0.280&  0.566&  0.459&  0.347&  0.301&  0.836&  0.993& 0.788\\
         CF100&  0.450&  0.428&  0.507&  0.557&   0.470&  0.438&  0.971&  0.801& 0.995\\
             \hline
    \end{tabular}
    }
    \label{tab: dataset similarity}
\end{table}
Evaluating the transferability of our \MyMethod inherently involves understanding the similarities between datasets, as this not only influences encoder transfer learning performance but also reflects the intrinsic characteristics of the data domains. 
Quantifying domain similarity is challenging, yet crucial for a comprehensive evaluation.
To address this, we adopt an approach inspired by previous work~\cite{li2021modeldiff}, utilizing cosine similarity as a metric to compare the features of input samples across different domains. 
We employ a widely-used feature extractor, a PyTorch pre-trained VGG-16 model, to extract latent features from pairs of data domains and calculate their cosine similarity.
The results, presented in Table~\ref{tab: dataset similarity}, corroborate the observations made in Section~\ref{sec: exp: target ntl} regarding the high similarity between the MT domain and the UP and EM domains—highlighted in bold within the table. 
This supports the notion that similarities in feature space significantly impact the transferability of both targeted and source-only \MyMethod.

\update{
\section{Training Cost of \MyMethod} \label{sec: complexity}
\begin{table}[h]
    \centering
    \update{
    \caption{Training time complexity of \MyMethod}    \label{tab: complexity}
    }
    \resizebox{0.9\linewidth}{!}{
    \begin{tabular}{l|llll|ccc}
    \hline
     \cellcolor{gray!20}Architecture & \cellcolor{gray!20}Level &  \cellcolor{gray!20}Source & \cellcolor{gray!20}Target & \cellcolor{gray!20}SIZE & \cellcolor{gray!20}DWS (s)  & \cellcolor{gray!20}DWU (s) & \cellcolor{gray!20}SC (s) \\
    \hline
    \cellcolor{koblue!20}ResNet-18    & \cellcolor{koblue!20}sup.  & \cellcolor{koblue!20}MT & \cellcolor{koblue!20}UP & \cellcolor{koblue!20}32 &\cellcolor{koblue!20}$0.771\pm0.147$  & \cellcolor{koblue!20}$5.85\pm0.021$ & \cellcolor{koblue!20}$2.33\pm1.59$ \\
    \cellcolor{darkblue!20}ResNet-18    & \cellcolor{darkblue!20}unsup./zero-shot  & \cellcolor{darkblue!20}MT & \cellcolor{darkblue!20}UP & \cellcolor{darkblue!20}32 & \cellcolor{darkblue!20}$1.03\pm0.086$ & \cellcolor{darkblue!20}$45.8\pm0.126$ & \cellcolor{darkblue!20}- \\
    \hline
    \cellcolor{koblue!20}VGG-11       & \cellcolor{koblue!20}sup. & \cellcolor{koblue!20}MT & \cellcolor{koblue!20}UP & \cellcolor{koblue!20}32 & \cellcolor{koblue!20}$0.821\pm0.275$  & \cellcolor{koblue!20}$8.66\pm0.039$  & \cellcolor{koblue!20}$8.75\pm9.11$\\
     \cellcolor{darkblue!20}VGG-11       &  \cellcolor{darkblue!20}unsup./zero-shot&  \cellcolor{darkblue!20}MT &  \cellcolor{darkblue!20}UP  &  \cellcolor{darkblue!20}32 &  \cellcolor{darkblue!20}$1.34\pm0.498$& \cellcolor{darkblue!20}$45.1\pm 0.122$ &  \cellcolor{darkblue!20}-\\
    \hline
    \cellcolor{koblue!20}ResNet-18    & \cellcolor{koblue!20}sup.  & \cellcolor{koblue!20}ImageNette & \cellcolor{koblue!20}Military  & \cellcolor{koblue!20}224 &  \cellcolor{koblue!20}$1.45\pm0.217$ & \cellcolor{koblue!20}$59.2\pm 2.23$  & \cellcolor{koblue!20}$53.1\pm10.2$\\
     \cellcolor{darkblue!20}ResNet-18    &  \cellcolor{darkblue!20}unsup./zero-shot &  \cellcolor{darkblue!20}ImageNette &  \cellcolor{darkblue!20}Military  &  \cellcolor{darkblue!20}224 & \cellcolor{darkblue!20}$1.82\pm0.286$ &  \cellcolor{darkblue!20}$455.2\pm1.03$ & \cellcolor{darkblue!20}- \\    
    \hline
    \end{tabular}
    }
\end{table}

In Table~\ref{tab: complexity}, we present the training time costs of supervised, unsupervised, and zero-shot \MyMethod on our platform equipped with an RTX TITAN GPU. Note the zero-shot \MyMethod has similar computational complexity as the unsupervised version, because they employ the same training strategy.
We choose to examine two representative source-target domain pairs: a smaller digits pair (MT and UP) with an input size of $32\times 32\times 3$, and a more complex image pair (ImageNette and Military) with an input size of $224\times 224\times 3$. Additionally, we assess the training costs associated with two different encoder architectures, ResNet-18 and VGG-11.
We break down the training complexity into the three key steps described in \MyMethod: domain-aware weight search (DWS), domain-aware weight updating (DWU), and self-challenging training (SC). The results demonstrate that the primary computational burden lies in the DWU step, and for supervised \MyMethod, the self-challenging phase is also costly, where the classifier must be retrained to enhance robustness against adversarial attacks.
Furthermore, we find that DWS and DWU in unsupervised and zero-shot \MyMethod incur higher computational costs compared to the supervised version. This is attributed to the contrastive learning-based loss function~\eqref{eq: unsupervised loss}.
When comparing different input sizes, higher-resolution images lead to substantially longer training time because of the larger feature maps to compute during the forward pass of the encoder, especially in the DWU process. Large models also require more training time. 
It is worth noting that
such training cost is a one-time expense, and there is no performance impact on the protected encoder during inference.
}

\section{Numerical results -- Supervised \MyMethod (ViT)}
\label{appendix: vit on simple datasets}
Following a similar setting in Section~\ref{sec: case-study}, we consider that the model provider aims to prevent the pre-trained ViT encoder on ImageNette from being transferred to a specified simple unauthorized domain. 
In particular, we evaluate \MyMethod on four target datasets: CF, ST, MT, and CF100.
To fit the input size of ViT, we resize the target input to $224\times224\times3$ and monitor the accuracy degradation on both the target datasets and ImageNet~\cite{howard2020fastai}, the source dataset.
During training, we follow the ViT fine-tuning instructions and use the last hidden state as the extracted feature space and connect to a single-dense-layer output classifier.
From results shown in Table~\ref{tab: exp: vit},
we find that \MyMethod reaches the non-transferability design goal --- reducing the ViT performance on the target domain by $65.8\%$ but keeping the accuracy on the source domain with a small accuracy drop of $2.15\%$.
Our experimental results demonstrate that \MyMethod is effective when applied to a large encoder pre-trained on a large amount of samples.

\begin{table}[h]
    \caption{\textbf{\MyMethod on ViT}--Source task (ImageNette~\cite{imagenette})}
    \centering
    \resizebox{0.9\linewidth}{!}{
    \begin{tabular}{c|c|c|c|c|c|c}
    \hline
    Target &  $Acc^{org}_{\mathcal{S}}$ & $Acc^{org}_{\mathcal{T}}$ & $Acc^{el}_{\mathcal{S}}$ & $Acc^{el}_{\mathcal{T}}$ & $Drop_{\mathcal{S}}$ & $Drop_{\mathcal{T}}$ \\
    \midrule
    CF       & \multirow{4}{*}{87.26}  & 74.99  & 86.83 & 36.70 &$0.49\%\downarrow$ & $51.9\%\downarrow$\\
    ST      &  & 73.50 & 86.80 & 10.53 & $0.53\%\downarrow$ &  $85.7\%\downarrow$\\
    MT     &  &  92.06 & 85.20  & 47.56 &  $2.36\%\downarrow$ & $48.3\%\downarrow$ \\
    CF100     &  &  53.31 & 82.62  & 12.07 &  $5.32\%\downarrow$ & $77.4\%\downarrow$ \\
    \hline
    \end{tabular}
    }
    \label{tab: exp: vit}
\end{table}


\section{Unsupervised \MyMethod on ResNet-18} \label{appendix: unsupervised resnet18}

\begin{table*}[h]
\centering
    \caption{\textbf{Unsupervised EncoderLock's performance on encoder (ResNet-18) transferability}}
    \resizebox{\linewidth}{!}{
    \begin{tabular}{c|c|c|c|c|c|c|c|c}
    \hline
         \makecell{Source \textbackslash{} Target}& MT & UP & SN & MM & SD & $\Delta W$ & $Drop_\mathcal{S}$ & $Drop_\mathcal{T}$\\
    \hline
    MT & $\cellcolor{gray!20} \mathbf{99.48\Rightarrow98.87}$ & $93.77\Rightarrow10.26$ & $41.47\Rightarrow11.27$ & $70.02\Rightarrow32.04$ & $72.48\Rightarrow16.17$ & $3.27$ \textperthousand  & $0.61\%\downarrow$ & $73.45\%\downarrow$ \\
    \hline
    UP & $95.10\Rightarrow37.66$ & \cellcolor{gray!20} $\mathbf{96.11\Rightarrow94.95}$ & $33.72\Rightarrow18.15$ & $55.79\Rightarrow18.67$ & $61.76\Rightarrow15.18$ & $2.91$ \textperthousand  & $1.21\%\downarrow$ & $62.13\%\downarrow$ \\
    \hline
    SN  & $94.65\Rightarrow22.62$ & $88.04\Rightarrow6.98$ & \cellcolor{gray!20} $\mathbf{91.06\Rightarrow89.55}$ & $66.52\Rightarrow13.92$ & $95.08\Rightarrow88.9$ & $1.41$ \textperthousand  & $1.66\%\downarrow$ & $63.43\%\downarrow$ \\
    \hline
    MM  & $98.82\Rightarrow17.94$ & $92.33\Rightarrow17.14$ & $48.18\Rightarrow18.18$ & \cellcolor{gray!20} $\mathbf{91.49\Rightarrow90.97}$ & $78.03\Rightarrow43.83$ & $1.19$ \textperthousand  & $0.57\%\downarrow$ & $67.34\%\downarrow$ \\
    \hline
    SD & $96.76\Rightarrow50.06$ & $91.68\Rightarrow21.72$ & $86.74\Rightarrow30.86$ & $68.56\Rightarrow17.68$ & \cellcolor{gray!20} $\mathbf{99.42\Rightarrow97.75}$ & $2.58$ \textperthousand  & $1.67\%\downarrow$ & $65.80\%\downarrow$\\
    \hline
    \end{tabular}
    }
    \label{tab: unsupervised encoderlock performance-resnet}
\end{table*}

In this section, we present the additional results using ResNet-18 for the unsupervised \MyMethod in Table~\ref{tab: unsupervised encoderlock performance-resnet}. 
The unsupervised \MyMethod also shows promising result in applicability authorization.

\update{
\section{Impact of the Synthetic Dataset Quality on Zero-shot \MyMethod} \label{appendix: zero-shot quality}
In addition to prompt relevance, the generation quality of the synthetic dataset significantly affects the performance of zero-shot \MyMethod. We measure the quality with two metrics: the noise level of the synthetic images and the generation quality of the diffusion model. To ensure a fair comparison, we use the same set of prompts (refined prompts from the prohibited domain) for zero-shot \MyMethod and select an encoder with the same degradation level on the authorized domain as reported in Section~\ref{exp: zero-shot encoderlock} (greater than $92\%$).
We report the \MyMethod performance in the format of ($acc_m^\mathcal{S}, acc_m^\mathcal{T}$). Note that the defense goal of \MyMethod is a higher $acc_m^\mathcal{S}$ and lower $acc_m^\mathcal{T}$.

\begin{figure}[h]
    \centering
    \includegraphics[width=0.85\linewidth]{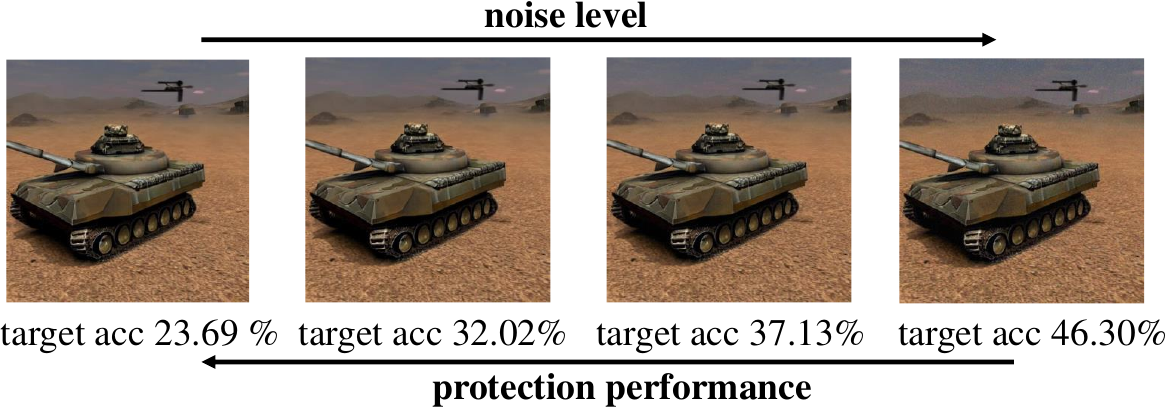}
    \caption{Zero-shot \MyMethod performance with different noise levels}
    \label{fig: noise}
\end{figure}

Fig.~\ref{fig: noise} shows an example image with different levels of Gaussian noise. Introducing random noise into the images reduces \MyMethod's ability to restrict the prohibited domain. We assess the impact of different noise levels, at $\sigma=1$, $\sigma=5$, and $\sigma=10$, respectively. The zero-shot \MyMethod performance degrades to ($91.90\%$, $32.02\%$),  ($91.85\%$, $37.13\%$), and ($92.20\%$, $46.30\%$), where the performance on the noise-free synthetic dataset is ($92.86\%$, $23.69\%$). 
High level of noise significantly reduces the synthetic quality, leading to poorer performance of the zero-shot \MyMethod.

\begin{figure}[h]
    \centering
    \includegraphics[width=0.85\linewidth]{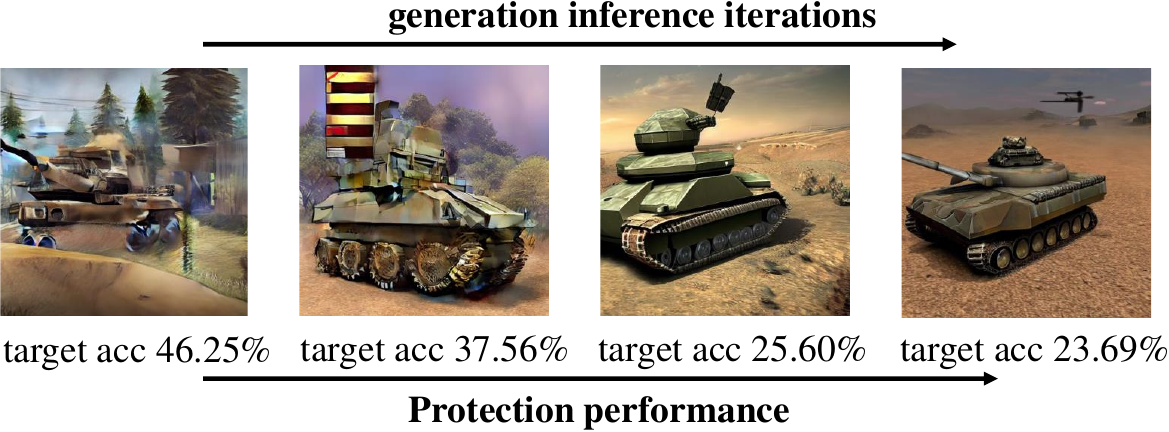}
    \caption{Zero-shot \MyMethod performance with different diffusion qualities}
    \label{fig: qualities}
\end{figure}
 
We regenerate the synthetic datasets with varying number of inference iterations in the stable diffusion model. The generation quality improves as we increase the iteration count from $5$, $10$, $20$, to $50$ (the value used in the original setting), as shown in 
Fig.~\ref{fig: qualities}. 
Consequently, the protection performance improves from ($92.71\%$, $46.25\%$) to ($92.25\%$, $37.56\%$), ($92.31\%$, $25.60\%$) and ($92.86\%$, $23.69\%$). 
A higher-quality synthetic dataset provides clearer potential features of the prohibited domain, thus enhancing the restriction performance. However, increasing the number of inference iterations in the generator leads to higher computational costs. Running 5 inference iterations achieves a speed of $8.2$ items/s, whereas 50 inference iterations reduce the speed to $0.8$ items/s.
}

\update{
\section{\MyMethod performance on Various Probing Data} \label{appendix: various data}
\begin{figure}[h]
    \centering
    \includegraphics[width=\linewidth]{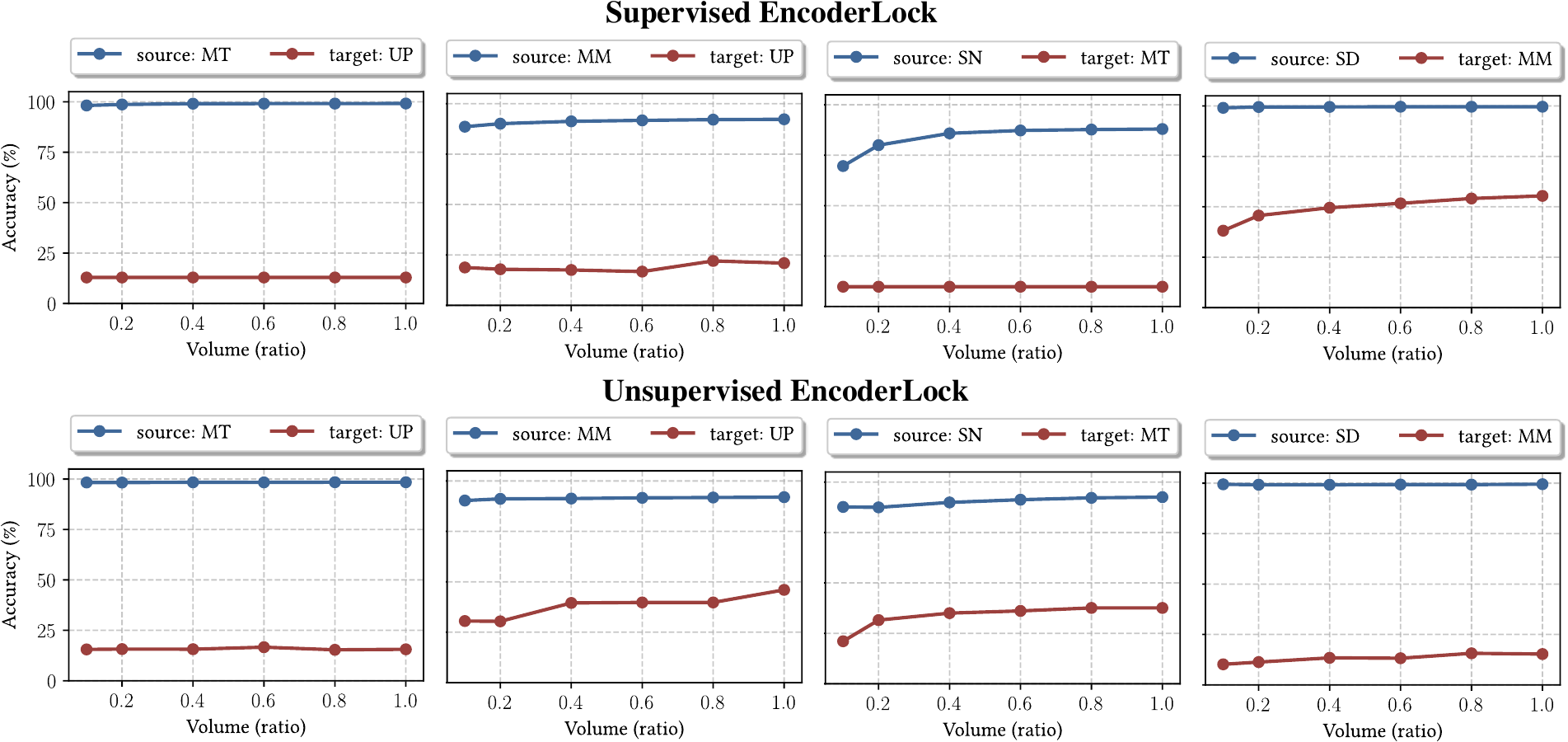}
    \caption{\MyMethod performance for various volumes of probing data.}
    \label{fig: data-volume}
\end{figure}
We evaluate the scenario where an attacker probes the encoder using varying amounts of probing data, ranging from $10\%$ to the entire dataset, in both supervised and unsupervised settings. The results are illustrated in Fig.~\ref{fig: data-volume}. 
Notably, for the prohibited target domain, the accuracy remains low even when the attacker utilizes the full prohibited dataset, while the protection slightly degrades with more data. 
For the authorized source domain, the accuracy remains consistently high even with a small amount of probing data.
These results demonstrate the robustness of both supervised and unsupervised versions of \MyMethod against malicious probing attempts.
}

\section{Compare with previous work} \label{appendix: comparison}
In this section, we present more comparison results with baselines similar to Table~\ref{tab: comparison with baseline methods}.
From Table~\ref{tab: comparison with baseline methods-appendix-1} to Table~\ref{tab: comparison with baseline methods-appendix-8}, we compare the baseline methods and the proposed supervised \MyMethod and unsupervised \MyMethod between different pairs of digit datasets.
The observation is similar to our conclusion in Section~\ref{exp: comparison}.
Specifically, under the condition of fine-tuning the downstream model fine-tuning, the proposed methods outperform baselines. 

\begin{table}[H]
    \centering
    \setlength{\fboxsep}{1pt} 
    \caption{Comparison the effectiveness of \MyMethod on the target domain and its performance on other domains with baselines~\cite{wang2022non, wang2023model}. In this table, \textbf{bold} text indicates the best performance, \underline{underlined}  denotes the second-best performance. The \colorbox{gray!20}{gray row} denotes original transfer accuracy.}
    \resizebox{0.9\linewidth}{!}{
    \begin{tabular}{c|c|c|c|c|c}
    \toprule
    \multirow{2}{*}{Methods \textbackslash{} Domain} & Source & Target  & \multicolumn{3}{c}{Other Domains} \\
    \cline{2-6}
    & MM & UP  & MT  & SN  & SD \\
    \hline
    \rowcolor{gray!20}
    Original Accuracy  & $94.2\%$ & $94.7\%$  & $98.9\%$ & $53.8\%$ & $85.9\%$ \\
    \hline
    NTL~\cite{wang2022non}   &  $81.6\%$ & $77.3\%$  & $97.9\%$ & $30.4\%$ & $62.1\%$ \\
    \hline
    CUTI~\cite{wang2023model}  &  $66.6\%$ & $90.6\%$  & $96.0\%$  & $\textbf{52.3}\%$  & $\textbf{82.0}\%$\\
    \hline
    \rowcolor{koblue!20}
    \textbf{Supervised \MyMethod} & $\textbf{93.5}\%$   & $\textbf{17.8}\%$   & $\underline{98.8\%}$    &  $39.3\%$  &  $69.1\%$\\ 
    \hline
    \rowcolor{darkblue!20}
    \textbf{Unsupervised \MyMethod} & $\underline{93.3\%}$   & $\underline{30.7\%}$   & $\textbf{98.9}\%$    &  $\underline{49.0\%}$  &  $\underline{78.4\%}$\\ 
    \hline
    \end{tabular}
    \label{tab: comparison with baseline methods}
    }
\end{table}

\begin{table}[H]
    \centering
    \setlength{\fboxsep}{1pt} 
    \caption{Comparison of \MyMethod with baselines}
    \resizebox{0.9\linewidth}{!}{
    \begin{tabular}{c|c|c|c|c|c}
    \toprule
    \multirow{2}{*}{Methods \textbackslash{} Domain} & Source & Target  & \multicolumn{3}{c}{Other Domains} \\
    \cline{2-6}
    & UP & MM  & MT  & SN  & SD \\
    \hline
    \rowcolor{gray!20}
    Original Accuracy  & $97.9\%$ & $65.2\%$  & $97.7\%$ & $58.2\%$ & $87.1\%$ \\
    \hline
    NTL~\cite{wang2022non}   &  $90.2\%$ & $\underline{12.8\%}$  & $\textbf{93.4\%}$  & $19.3\%$  & $33.8\%$ \\
    \hline
    CUTI~\cite{wang2023model}  &  $93.9\%$ & $33.7\%$  & $\underline{93.6\%}$ & $\underline{31.1\%}$ & $\textbf{78.9\%}$\\
    \hline
    \rowcolor{koblue!20}
    \textbf{Supervised \MyMethod} & $\underline{96.9}\%$   & $15.6\%$   & $81.2\%$    &  $\textbf{31.2\%}$  &  $\underline{60.9\%}$\\ 
    \hline
    \rowcolor{darkblue!20}
    \textbf{Unsupervised \MyMethod} & $\textbf{96.7\%}$   & $\textbf{9.59\%}$   & $60.5\%$    &  $22.6\%$  &  $31.9\%$\\ 
    \hline
    \end{tabular}
    \label{tab: comparison with baseline methods-appendix-1}
    }
\end{table}

\begin{table}[H]
    \centering
    \setlength{\fboxsep}{1pt} 
    \caption{Comparison of \MyMethod with baselines}
    \resizebox{0.9\linewidth}{!}{
    \begin{tabular}{c|c|c|c|c|c}
    \toprule
    \multirow{2}{*}{Methods \textbackslash{} Domain} & Source & Target  & \multicolumn{3}{c}{Other Domains} \\
    \cline{2-6}
    & MT & UP  & MM  & SN  & SD \\
    \hline
    \rowcolor{gray!20}
    Original Accuracy  & $99.5\%$ & $96.4\%$  & $68.2\%$ & $43.7\%$ & $69.6\%$ \\
    \hline
    NTL~\cite{wang2022non}   &  $97.4\%$ & $78.1\%$  & $37.4\%$ & $19.8\%$ & $38.6\%$ \\
    \hline
    CUTI~\cite{wang2023model}  &  $97.9\%$ & $93.6\%$  & $\textbf{66.5\%}$  & $\textbf{48.4}\%$  & $\textbf{86.5}\%$\\
    \hline
    \rowcolor{koblue!20}
    \textbf{Supervised \MyMethod} & $\textbf{99.3}\%$   & $\textbf{8.47}\%$   & $25.3\%$    &  $19.6\%$  &  $14.5\%$\\ 
    \hline
    \rowcolor{darkblue!20}
    \textbf{Unsupervised \MyMethod} & $\underline{99.1\%}$   & $\underline{16.8\%}$   & $\underline{49.3}\%$    &  $\underline{25.4\%}$  &  $\underline{43.1\%}$\\ 
    \hline
    \end{tabular}
    \label{tab: comparison with baseline methods-appendix-2}
    }
\end{table}

\begin{table}[H]
    \centering
    \setlength{\fboxsep}{1pt} 
    \caption{Comparison of \MyMethod with baselines}
    \resizebox{0.9\linewidth}{!}{
    \begin{tabular}{c|c|c|c|c|c}
    \toprule
    \multirow{2}{*}{Methods \textbackslash{} Domain} & Source & Target  & \multicolumn{3}{c}{Other Domains} \\
    \cline{2-6}
    & SN & UP  & MT  & MM  & SD \\
    \hline
    \rowcolor{gray!20}
    Original Accuracy  & $94.0\%$ & $92.7\%$  & $95.3\%$ & $71.5\%$ & $97.0\%$ \\
    \hline
    NTL~\cite{wang2022non}   &  $78.0\%$ & $88.4 \%$  & $ \textbf{96.3\%}$ & $\underline{72.3\%}$ & $\underline{92.6\%}$ \\
    \hline
    CUTI~\cite{wang2023model}  &  $62.1\%$ & $91.3\%$  & $\underline{95.5\%}$  & $67.0\%$  & $82.2\%$\\
    \hline
    \rowcolor{koblue!20}
    \textbf{Supervised \MyMethod} & $\underline{87.3\%}$   & $\underline{19.7}\%$   & $12.0\%$    &  $51.9\%$  &  $84.8\%$\\ 
    \hline
    \rowcolor{darkblue!20}
    \textbf{Unsupervised \MyMethod} & $\textbf{94.7\%}$   & $\textbf{17.1\%}$   & $94.2\%$    &  $\textbf{70.3\%}$  &  $\textbf{96.9\%}$\\ 
    \hline
    \end{tabular}
    \label{tab: comparison with baseline methods-appendix-3}
    }
\end{table}

\begin{table}[H]
    \centering
    \setlength{\fboxsep}{1pt} 

    \caption{Comparison of \MyMethod with baselines}
    \resizebox{0.9\linewidth}{!}{
    \begin{tabular}{c|c|c|c|c|c}
    \toprule
    \multirow{2}{*}{Methods \textbackslash{} Domain} & Source & Target  & \multicolumn{3}{c}{Other Domains} \\
    \cline{2-6}
    & SD & MT  & UP  & SN  & MM \\
    \hline
    \rowcolor{gray!20}
    Original Accuracy  & $99.8\%$ & $97.1\%$  & $93.6\%$ & $53.8\%$ & $71.5\%$ \\
    \hline
    NTL~\cite{wang2022non}   &  $95.9\%$ & $90.6\%$  & $89.6\%$ & $57.0\%$ & $62.0\%$ \\
    \hline
    CUTI~\cite{wang2023model}  &  $87.1\%$ & $96.0\%$  & $88.7\%$  & $61.9\%$  & $\textbf{64.3}\%$\\
    \hline
    \rowcolor{koblue!20}
    \textbf{Supervised \MyMethod} & $\textbf{99.6}\%$   & $\textbf{11.9}\%$   & $\underline{90.4\%}$    &  $\textbf{88.9}\%$  &  $50.6\%$\\ 
    \hline
    \rowcolor{darkblue!20}
    \textbf{Unsupervised \MyMethod} & $\underline{99.5\%}$   & $\underline{76.6\%}$   & $\textbf{90.9}\%$    &  $\underline{87.9\%}$  &  $\underline{62.3\%}$\\ 
    \hline
    \end{tabular}
    \label{tab: comparison with baseline methods-appendix-4}
    }
\end{table}

\begin{table}[H]
    \centering
    \setlength{\fboxsep}{1pt} 
    \caption{Comparison of \MyMethod with baselines}
    \resizebox{0.9\linewidth}{!}{
    \begin{tabular}{c|c|c|c|c|c}
    \toprule
    \multirow{2}{*}{Methods \textbackslash{} Domain} & Source & Target  & \multicolumn{3}{c}{Other Domains} \\
    \cline{2-6}
    & UP & SD  & MT  & SN  & MM \\
    \hline
    \rowcolor{gray!20}
    Original Accuracy  & $97.9\%$ & $87.1\%$  & $97.7\%$ & $58.2\%$ & $65.2\%$ \\
    \hline
    NTL~\cite{wang2022non}   &  $89.2\%$ & $28.2\%$  & $89.7\%$  & $18.7\%$  & $17.1\%$ \\
    \hline
    CUTI~\cite{wang2023model}  &  $92.8\%$ & $77.7\%$  & $\underline{95.0\%}$ & $\underline{35.1\%}$ & $37.3\%$\\
    \hline
    \rowcolor{koblue!20}
    \textbf{Supervised \MyMethod} & $\textbf{96.8}\%$   & $\underline{16.9\%}$   & $\textbf{96.8\%}$    &  $\textbf{45.7\%}$  &  $\textbf{57.5\%}$\\ 
    \hline
    \rowcolor{darkblue!20}
    \textbf{Unsupervised \MyMethod} & $\underline{96.7\%}$   & $\textbf{12.7\%}$   & $\textbf{88.0}\%$    &  $19.7\%$  &  $\underline{42.3\%}$\\ 
    \hline
    \end{tabular}
    \label{tab: comparison with baseline methods-appendix-5}
    }
\end{table}

\begin{table}[H]
    \centering
    \setlength{\fboxsep}{1pt} 
    \caption{Comparison of \MyMethod with baselines}
    \resizebox{0.9\linewidth}{!}{
    \begin{tabular}{c|c|c|c|c|c}
    \toprule
    \multirow{2}{*}{Methods \textbackslash{} Domain} & Source & Target  & \multicolumn{3}{c}{Other Domains} \\
    \cline{2-6}
    & MT & MM  & UP  & SN  & SD \\
    \hline
    \rowcolor{gray!20}
    Original Accuracy  & $99.5\%$ & $68.2\%$  & $96.4\%$ & $43.7\%$ & $69.6\%$ \\
    \hline
    NTL~\cite{wang2022non}   &  $97.6\%$ & $33.7\%$  & $79.9\%$ & $19.8\%$ & $33.4\%$ \\
    \hline
    CUTI~\cite{wang2023model}  &  $98.1\%$ & $56.1\%$  & $\textbf{93.2\%}$  & $\textbf{34.4}\%$  & $\textbf{84.9}\%$\\
    \hline
    \rowcolor{koblue!20}
    \textbf{Supervised \MyMethod} & $\textbf{99.3}\%$   & $\underline{23.2}\%$   & $69.3\%$    &  $19.6\%$  &  $31.4\%$\\ 
    \hline
    \rowcolor{darkblue!20}
    \textbf{Unsupervised \MyMethod} & $\underline{99.2\%}$   & $\textbf{12.3\%}$   & $\underline{87.1}\%$    &  $\underline{26.1\%}$  &  $\underline{43.6\%}$\\ 
    \hline
    \end{tabular}
    \label{tab: comparison with baseline methods-appendix-6}
    }
\end{table}

\begin{table}[H]
    \centering
    \setlength{\fboxsep}{1pt} 
    \caption{Comparison of \MyMethod with baselines}
    \resizebox{0.9\linewidth}{!}{
    \begin{tabular}{c|c|c|c|c|c}
    \toprule
    \multirow{2}{*}{Methods \textbackslash{} Domain} & Source & Target  & \multicolumn{3}{c}{Other Domains} \\
    \cline{2-6}
    & SN & MM  & MT  & UP  & SD \\
    \hline
    \rowcolor{gray!20}
    Original Accuracy  & $94.0\%$ & $71.5\%$  & $95.3\%$ & $92.7\%$ & $97.0\%$ \\
    \hline
    NTL~\cite{wang2022non}   &  $77.8\%$ & $72.6 \%$  & $\textbf{96.8\%}$ & $89.8\%$ & $92.2\%$ \\
    \hline
    CUTI~\cite{wang2023model}  &  $65.4\%$ & $61.5\%$  & $\underline{95.5\%}$  & $\textbf{89.7}\%$  & $87.9\%$\\
    \hline
    \rowcolor{koblue!20}
    \textbf{Supervised \MyMethod} & $\underline{91.8}\%$   & $\textbf{14.6}\%$   & ${17.3\%}$    &  $\underline{93.1\%}$  &  $\underline{91.8\%}$\\ 
    \hline
    \rowcolor{darkblue!20}
    \textbf{Unsupervised \MyMethod} & $\textbf{94.6\%}$   & $\underline{50.8\%}$   & ${93.3}\%$    &  $\textbf{93.8\%}$  &  $\textbf{97.1\%}$\\ 
    \hline
    \end{tabular}
    \label{tab: comparison with baseline methods-appendix-7}
    }
\end{table}

\begin{table}[H]
    \centering
    \setlength{\fboxsep}{1pt} 
    \caption{Comparison of \MyMethod with baselines}
    \resizebox{0.9\linewidth}{!}{
    \begin{tabular}{c|c|c|c|c|c}
    \toprule
    \multirow{2}{*}{Methods \textbackslash{} Domain} & Source & Target  & \multicolumn{3}{c}{Other Domains} \\
    \cline{2-6}
    & SD & MM  & UP  & SN  & MT \\
    \hline
    \rowcolor{gray!20}
    Original Accuracy  & $99.8\%$ & $71.5\%$  & $93.6\%$ & $53.8\%$ & $97.1\%$ \\
    \hline
    NTL~\cite{wang2022non}   &  $89.8\%$ & $56.1\%$  & $90.5\%$ & $51.4\%$ & $96.1\%$ \\
    \hline
    CUTI~\cite{wang2023model}  &  $90.7\%$ & $66.1\%$  & $\underline{91.5\%}$  & ${67.1}\%$  & $\underline{96.8}\%$\\
    \hline
    \rowcolor{koblue!20}
    \textbf{Supervised \MyMethod} & $\textbf{99.8}\%$   & $\underline{40.9}\%$   & $\textbf{94.2\%}$    &  $\underline{90.2\%}$  &  $\textbf{97.2\%}$\\ 
    \hline
    \rowcolor{darkblue!20}
    \textbf{Unsupervised \MyMethod} & $\underline{99.5\%}$   & $\textbf{27.2\%}$   & ${88.7}\%$    &  $\textbf{99.6\%}$  &  ${88.8\%}$\\ 
    \hline
    \end{tabular}
    \label{tab: comparison with baseline methods-appendix-8}
    }
\end{table}


\update{
\section{Security Analysis--Train-from-scratch accuracy} \label{appendix: security analysis}
\begin{figure}[h]
    \centering
    \includegraphics[width=\linewidth]{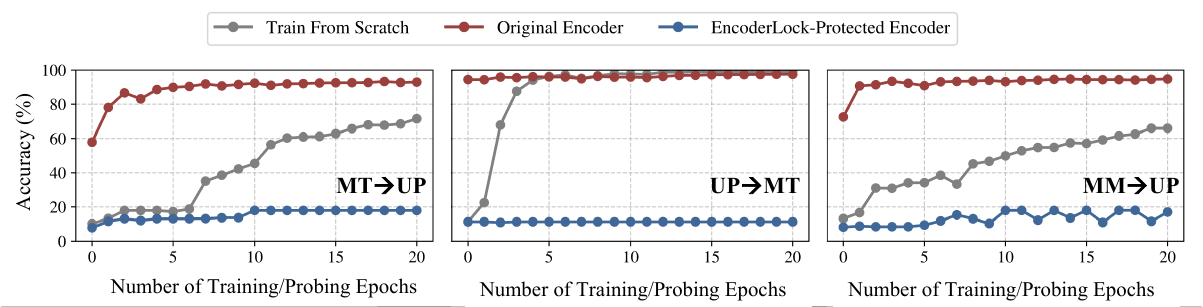}
    \caption{\MyMethod's Performance Versus Train-from-scratch}
    \label{fig: tfs}
\end{figure}

Here we apply the security assessment definition in Section~\ref{sec: security analysis} on three example pairs of domains: \textit{MT to UP}, \textit{UP to MT}, and \textit{MM to UP} for the supervised \MyMethod on VGG-11. Their accuracy drops on the authorized domain are $0.07\%$, $0.25\%$, and $0.17\%$, respectively, all below the accuracy drop constraint ($\epsilon$ = $2\%$).
Fig.~\ref{fig: tfs} shows the probing performance of the \MyMethod-protected and unprotected encoders on the prohibited domain, compared to the accuracy of the ``train-from-scratch" model. It demonstrates that starting from the \MyMethod-protected encoder allows the model to achieve lower accuracy and faster convergence on the prohibited domain, than a model trained from scratch. 
Therefore, 
the protected encoder can be considered \textbf{SECURE}, as an attacker would have no motivation to perform malicious probing.
 By contrast, the original encoder is \textbf{NOT SECURE}, as the accuracy ({\color{red} red}) is always higher than train-from-scratch accuracy ({\color{gray} gray}) as shown in Fig.~\ref{fig: tfs}. 
}

\section{\MyMethod GradCAM on Admissible Domains} \label{append: interpretation admissible}

\begin{figure}[h]
    \centering
    \includegraphics[width=\linewidth]{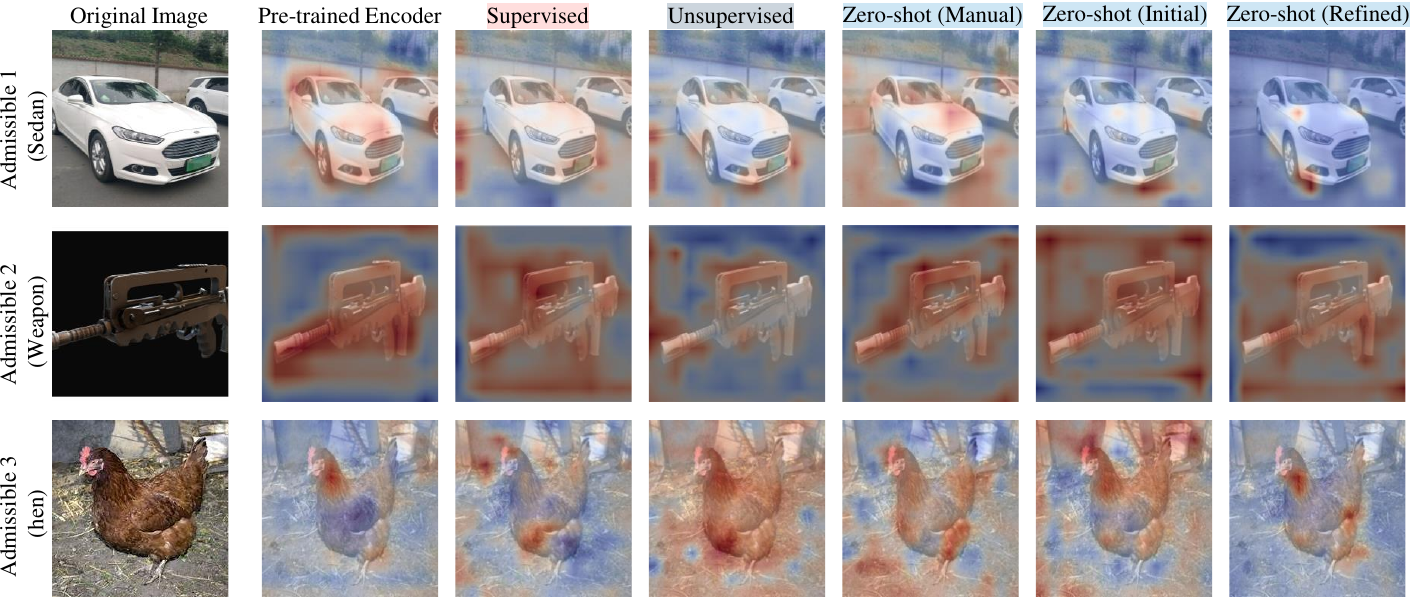}
    \caption{\textbf{Interpretation of Different \MyMethod using GradCAM~\cite{selvaraju2017grad}}--the {\color{red} red} parts highlight the focus of encoder to make decisions.}
    \label{fig: visual interpretation-2}
\end{figure}
In this section, we present the additional results to visualize the admissible domains with GradCAM in Fig.~\ref{fig: visual interpretation-2}. 

\section{Generated prompts \& images} 
\label{appendix: synthetic images}

\begin{figure}[h]
    \centering
    \includegraphics[width=\linewidth]{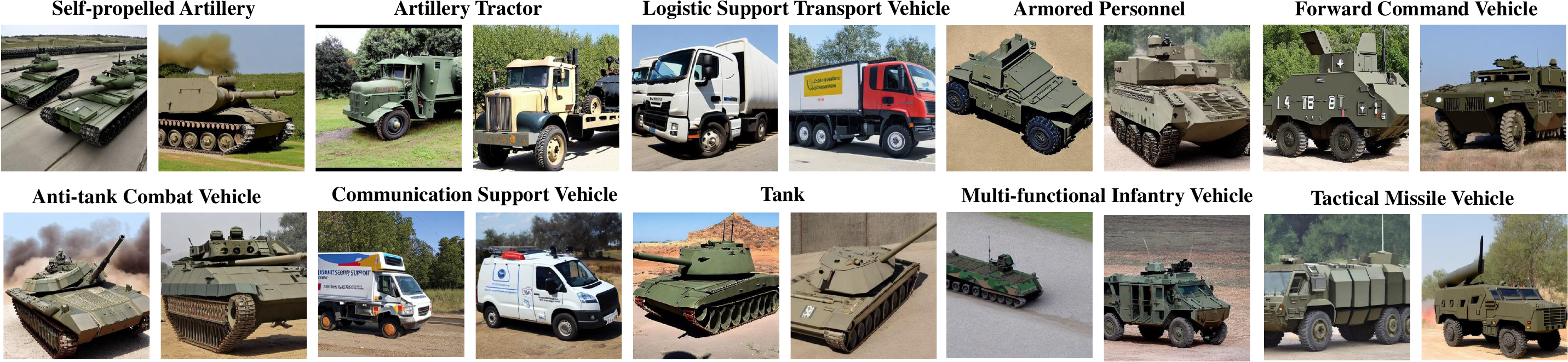}
    \caption{\textbf{Manual Prompts} and Generated Synthetic Dataset}
    \label{fig:refined-prompts-1}
\end{figure}

\begin{figure}[h]
    \centering
    \includegraphics[width=\linewidth]{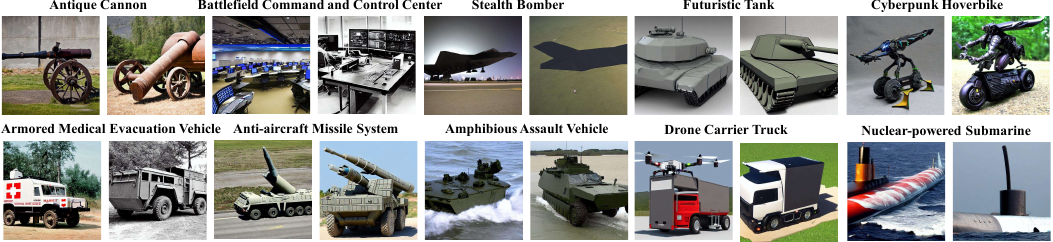}
    \caption{\textbf{AI agent Initial Prompts} and Generated Synthetic Dataset}
    \label{fig: initial-prompts}
\end{figure}

The (refined) prompts for generating synthetic datasets in zero-shot \MyMethod.
The theme is \textbf{military vehicles}.
\newpage
\noindent\textbf{Manual Prompts. See Figure~\ref{fig:refined-prompts-1}}
\vspace{-4mm}
\begin{multicols}{2}
\scriptsize
\begin{itemize}
    \item Armored Personnel Carrier
    \item Anti-tank Combat Vehicle
    \item Tactical Missile Vehicle
    \item Forward Command Vehicle
    \item Communication Support Vehicle
    \item Artillery Tractor
    \item Logistic Support Transport Vehicle
    \item Tank
    \item Self-propelled Artillery
    \item Multi-functional Infantry Vehicle
\end{itemize}
\end{multicols}
\vspace{-4mm}
\noindent\textbf{Initial Prompts: The synthetic images shown in Figure~\ref{fig: initial-prompts}}
\vspace{-4mm}
\begin{multicols}{2}
\scriptsize
\begin{itemize}
    \item futuristic tank, stealth design
    \item antique cannon, ceremonial use 
    \item amphibious assault vehicle, coastal operations
    \item drone carrier truck, mobile base  
    \item armored medical evacuation vehicle, red cross  
    \item cyberpunk hoverbike, scout unit  
    \item nuclear-powered submarine, deep-sea exploration 
    \item stealth bomber, night operation  
    \item battlefield command and control center, high-tech  
    \item anti-aircraft missile system, mobile defense
\end{itemize}
\end{multicols}

\begin{figure}[h]
    \centering
    \includegraphics[width=\linewidth]{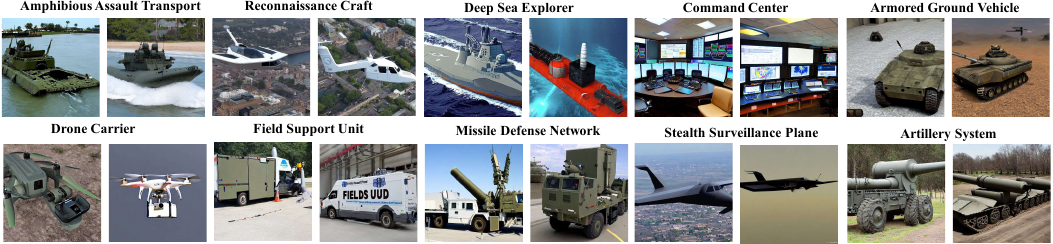}
    \caption{\textbf{AI agent Refined Prompts} and Generated Synthetic Dataset}
    \label{fig:refined-prompts}
\end{figure}

\vspace{-4mm}
\noindent\textbf{Refined Prompts: The synthetic image shown in Figure~\ref{fig:refined-prompts}}
\vspace{-4mm}
\begin{multicols}{2}
\scriptsize
\begin{itemize}

    \item Armored Ground Vehicle, Modern Combat  
    \item Artillery System, Classic Aesthetics  
    \item Amphibious Assault Transport  
    \item Drone Carrier, Tactical  
    \item Field Support Unit, Healthcare  
    \item Reconnaissance Craft, Urban Aerial  
    \item Deep Sea Explorer, Nuclear Propulsion  
    \item Stealth Surveillance Plane  
    \item Command Center, High-Tech  
    \item Missile Defense Network, Mobile
\end{itemize}
\end{multicols}

\end{document}